\makeatletter
\def\input@path{{styles/}}
\makeatother
\documentclass[11pt]{article}
\def\UseBibLatex{1}

\makeatletter
\def\input@path{{styles/}}
\makeatother

\providecommand{\BibLatexMode}[1]{}
\providecommand{\BibTexMode}[1]{}

\renewcommand{\BibLatexMode}[1]{#1}
\renewcommand{\BibTexMode}[1]{}

\ifx\UseBibLatex\undefined%
  \renewcommand{\BibLatexMode}[1]{}
  \renewcommand{\BibTexMode}[1]{#1}
\fi

\BibLatexMode{%
   \usepackage[bibencoding=utf8,style=alphabetic,backend=biber]{biblatex}%
   \usepackage{sariel_biblatex}%
}

\usepackage[in]{fullpage}%
\usepackage{amsmath}%
\usepackage{amssymb}%
\usepackage[table]{xcolor}%

\usepackage[amsmath,thmmarks]{ntheorem}%

\usepackage{microtype}
\usepackage{xcolor}%
\usepackage{mleftright}%
\usepackage{xspace}%
\usepackage{graphicx}
\usepackage{hyperref}%
\usepackage[inline]{enumitem} %
\usepackage{hyperref}%

\hypersetup{%
      unicode,
      breaklinks,%
      colorlinks=true,%
      urlcolor=[rgb]{0.25,0.0,0.0},%
      linkcolor=[rgb]{0.5,0.0,0.0},%
      citecolor=[rgb]{0,0.2,0.445},%
      filecolor=[rgb]{0,0,0.4},
      anchorcolor=[rgb]={0.0,0.1,0.2}%
}

\usepackage{titlesec}%
\titlelabel{\thetitle. }%

\usepackage[amsmath,thmmarks]{ntheorem}%
\usepackage{pgffor} %

\theoremseparator{.}%

\theoremstyle{plain}%
\theoremheaderfont{\normalfont\bfseries}
\theorembodyfont{\itshape}
\newtheorem{theorem}{Theorem}[section]

\foreach \env/\display in {   
   lemma/Lemma, conjecture/Conjecture, corollary/Corollary, claim/Claim,  invariant/Invariant,
   proposition/Proposition, prop/Proposition, openproblem/Open Problem%
}{ \edef\temp{{\noexpand\newtheorem{\env}[theorem]{\display}}} \temp }

\theoremseparator{}%

\theoremseparator{.}%

\definecolor{darkslate}{rgb}{0.2, 0.2, 0.2} %

\theoremseparator{.}%
\theoremheaderfont{\sffamily\bfseries\color{darkslate}}
\theorembodyfont{\upshape}%
\foreach \env/\display in {%
   defn/Definition, definition/Definition, example/Example, exercise/Exercise, xca/Exercise, problem/Problem,
   question/Question,assumption/Assumption%
}{\edef\temp{{\noexpand\newtheorem{\env}[theorem]{\display}}} \temp}

\theoremstyle{plain}%
\theoremseparator{.}%
\theoremheaderfont{\sffamily}%
\theorembodyfont{\upshape}%
\newtheorem*{remark:unnumbered}[theorem]{Remark}%

\foreach \env/\display in {
   remark/Remark, note/Note, fact/Fact, observation/Observation
}{ \edef\temp{{\noexpand\newtheorem{\env}[theorem]{\display}}} \temp }

\newcommand{\myqedsymbol}{\ensuremath{\blacksquare}}

\theoremheaderfont{\itshape}%
\theorembodyfont{\upshape}%
\theoremstyle{nonumberplain}%
\theoremseparator{}%
\theoremsymbol{\myqedsymbol}%
\newtheorem{proof}{Proof:}%
\newtheorem{proofof}{Proof of\!}%

\definecolor{blue25emph}{rgb}{0, 0, 11}
\definecolor{almostblack}{rgb}{0, 0, 0.3}

\providecommand{\emphi}[1]{}
\renewcommand{\emphi}[1]{\textcolor{blue25emph}{\textbf{\emph{#1}}}}

\providecommand{\emphw}[1]{}%
\renewcommand{\emphw}[1]{{\textcolor{almostblack}{\emph{#1}}}}%

\providecommand{\emphOnly}[1]{}%
\renewcommand{\emphOnly}[1]{\emph{\textcolor{blue25emph}{\textbf{#1}}}}

\newcommand{\SarielThanks}[1]{%
   \thanks{%
      School of Computing and Data Science; %
      University of Illinois; %
      201 N. Goodwin Avenue; %
      Urbana, IL, 61801, USA; %
      \href{mailto:spam@illinois.edu}{sariel@illinois.edu}; %
      \url{http://sarielhp.org/}.%
   #1%
   }%
}

\newcommand{\HLink}[2]{\hyperref[#2]{#1~\ref*{#2}}}
\newcommand{\HLinkSuffix}[3]{\hyperref[#2]{#1\ref*{#2}{#3}}}

\newcommand{\figlab}[1]{\label{fig:#1}}
\newcommand{\figref}[1]{\HLink{Figure}{fig:#1}}

\newcommand{\thmlab}[1]{{\label{theo:#1}}}
\newcommand{\thmref}[1]{\HLink{Theorem}{theo:#1}}

\newcommand{\corlab}[1]{\label{cor:#1}}
\newcommand{\corref}[1]{\HLink{Corollary}{cor:#1}}%

\newcommand{\lemlab}[1]{\label{lemma:#1}}
\newcommand{\lemref}[1]{\HLink{Lemma}{lemma:#1}}%

\newcommand{\seclab}[1]{\label{sec:#1}}
\newcommand{\secref}[1]{\HLink{Section}{sec:#1}}

\providecommand{\deflab}[1]{\label{def:#1}}
\newcommand{\defref}[1]{\HLink{Definition}{def:#1}}
\newcommand{\defrefY}[2]{\hyperref[def:#1]{#2}}

\providecommand{\eqlab}[1]{}%
\renewcommand{\eqlab}[1]{\label{equation:#1}}
\newcommand{\Eqref}[1]{\HLinkSuffix{Eq.~(}{equation:#1}{)}}

\providecommand{\remove}[1]{}%
\newcommand{\Set}[2]{\left\{ #1 \;\middle\vert\; #2 \right\}}

\newcommand{\pth}[1]{\mleft(#1\mright)}%

\newcommand{\cardin}[1]{\left\lvert {#1} \right\rvert}%

\newlist{compactenumA}{enumerate}{5}%
\setlist[compactenumA]{itemsep=-0.5ex,topsep=0.5ex,partopsep=1ex,parsep=1ex,%
   label=(\Alph*)}%

\newlist{compactenuma}{enumerate}{5}%
\setlist[compactenuma]{itemsep=-0.5ex,topsep=0.5ex,partopsep=1ex,parsep=1ex,%
   label=(\alph*)}%

\newlist{compactenumI}{enumerate}{5}%
\setlist[compactenumI]{itemsep=-0.5ex,topsep=0.5ex,partopsep=1ex,parsep=1ex,%
   label=(\Roman*)}%

\newlist{compactenumi}{enumerate}{5}%
\setlist[compactenumi]{itemsep=-0.5ex,topsep=0.5ex,partopsep=1ex,parsep=1ex,%
   label=(\roman*)}%

\newlist{compactitem}{itemize}{5}%
\setlist[compactitem]{itemsep=-0.5ex,topsep=0.5ex,partopsep=1ex,parsep=1ex,%
   label=\ensuremath{\bullet}}%

\newcommand{\etal}{\textit{et~al.}\xspace}

\numberwithin{figure}{section}%
\numberwithin{table}{section}%
\numberwithin{equation}{section}%

\usepackage[most]{tcolorbox}
\usepackage{marginnote}

\newcommand{\NewAuthor}[4]{%
    \expandafter\newcommand\csname #1margin\endcsname[1]{%
        \marginnote{\begin{tcolorbox}[
            enhanced, sharp corners, boxrule=0.5pt,
            colback=#2!10, colframe=#2!80!black,
            fontupper=\tiny\sffamily, width=\marginparwidth,
            left=2pt, right=2pt, top=2pt, bottom=2pt
        ] \textbf{#3:} ##1 \end{tcolorbox}}%
    }

    \newenvironment{#1block}[1]
    {\begin{tcolorbox}[
        colback=#2!5, colframe=#2!75!black,
        fonttitle=\bfseries, title=#4's Note: ##1,
        breakable, enhanced
    ]}
     {\end{tcolorbox}}
 \expandafter\newcommand\csname #4\endcsname[2][]{%
    \begin{#1block}{##1}
        ##2
    \end{#1block}
 }

 \expandafter\newcommand\csname #1\endcsname[1]{%
    {\color{#2} \textbf{***} \textbf{#3:} ##1 \textbf{***}}
 }
  \expandafter\newcommand\csname m#1\endcsname[1]{%
     \begin{#1margin}
         ##1 
     \end{#1margin}
  }
}

\AtBeginDocument{%
   \setlength{\marginparwidth}{\dimexpr\paperwidth - 1in - \oddsidemargin
      - \textwidth - \marginparsep - 10pt\relax}%
}

\NewAuthor{sariel}{blue}{SH}{Sariel}

\usepackage{mathtools}%
\usepackage{booktabs}%
\usepackage{placeins}%

\newcommand{\DrorThanks}[1]{%
   \thanks{%
      Google Research; %
      \href{mailto:aigerd@google.com}{aigerd@google.com}. %
      #1%
   }%
}

\theoremheaderfont{\itshape}%
\theorembodyfont{\upshape}%
\theoremstyle{nonumberplain}%
\theoremseparator{}%
\theoremsymbol{\myqedsymbol}%
\newtheorem{proofidea}{Proof idea:}%

\providecommand{\HLinkY}[2]{\hyperref[#2]{#1}}

\newcommand{\tablab}[1]{\label{tab:#1}}
\newcommand{\tabref}[1]{\HLink{Table}{tab:#1}}

\newcommand{\proplab}[1]{\label{prop:#1}}
\newcommand{\propref}[1]{\HLink{Proposition}{prop:#1}}%

\newcommand{\condlab}[1]{\label{cond:#1}}
\NewDocumentCommand{\condref}{O{Condition} m}{%
    \hyperref[cond:#2]{\IfBlankF{#1}{#1~}\textup{(\ref*{cond:#2})}}}

\newcommand{\clmlab}[1]{\label{claim:#1}}
\newcommand{\clmref}[1]{\HLink{Claim}{claim:#1}}%

\newcommand{\obslab}[1]{\label{obs:#1}}
\newcommand{\obsref}[1]{\HLink{Observation}{obs:#1}}%

\newcommand{\apndlab}[1]{\label{apnd:#1}}
\newcommand{\apndref}[1]{\HLink{Appendix}{apnd:#1}}

\newcommand{\abs}[1]{\left\lvert {#1} \right\rvert}%

\newenvironment{conditions}[1]
    {\begin{compactenumi}[label=\textup{(#1\arabic*)},ref=#1\arabic*]}
    {\end{compactenumi}}

\providecommand{\R}{\mathbb{R}}
\providecommand{\pth}[1]{\left(#1\right)}
\newcommand{\Z}{\mathbb{Z}}
\newcommand{\W}{\mathcal{W}}
\newcommand{\orient}{\operatorname{orient}}
\newcommand{\witnessfamily}[2]{\W(#1,#2)}
\newcommand{\oppcone}[3]{U(#1,#2,#3)}
\newcommand{\cone}{\operatorname{cone}}
\newcommand{\CHX}[1]{\mathcal{CH}\pth{#1}}
\newcommand{\vis}{\operatorname{vis}}
\newcommand{\inte}{\operatorname{int}}
\newcommand{\sgn}{\operatorname{sgn}}
\newcommand{\vcdim}{\operatorname{VC-dim}}
\newcommand{\vcdimconv}{\operatorname{VC-dim}_{\mathrm{convex}}}
\newcommand{\dist}{\operatorname{dist}}
\newcommand{\C}{\mathcal{C}}
\newcommand{\mD}{\mathsf{D}}
\newcommand{\mF}{\mathsf{F}}
\newcommand{\mH}{\mathsf{H}}
\newcommand{\mK}{\mathsf{K}}

\newcommand{\mL}{\mathsf{L}}
\newcommand{\mM}{\mathsf{M}}
\newcommand{\mN}{\mathsf{N}}
\newcommand{\mS}{\mathsf{S}}
\newcommand{\mT}{\mathsf{T}}
\newcommand{\mV}{\mathsf{V}}
\newcommand{\mX}{\mathsf{X}}
\newcommand{\mo}{\mathsf{o}}
\newcommand{\mr}{\mathsf{r}}
\newcommand{\cD}{\mathcal{D}}

\newcommand{\normorient}{\orient_{\mathrm{norm}}}

\newcommand{\Sin}{\mS_{\mathsf{in}}}
\newcommand{\missw}[1]{w_{\neg #1}}

\BibLatexMode{%
   \addbibresource{vc_dim_6.bib}
}

\begin{document}

\title{The VC-Dimension of Visibility in Simple Polygons Is Six}

\author{%
   Dror Aiger\DrorThanks{}%
   \and
   Sariel Har-Peled\SarielThanks{ Work on this paper
      was partially supported by NSF AF award
      CCF-2317241.
   }}

\date{\today}

\maketitle

\begin{abstract}
    Six targets in a simple polygon can be placed so that, for every
    subset of the targets, some point in the polygon sees precisely that
    subset---that is, the polygon \emph{shatters} the six targets. We
    prove that no simple polygon shatters seven targets. Together with
    Valtr's six-target construction, this determines the maximum
    visibility VC-dimension of a simple polygon to be six (even when the
    targets must lie in strictly convex position) and improves the
    previous upper bound of $14$ due to Gilbers and Klein. To rule out a
    shattered set of seven targets, we classify its convex-hull
    structure into $74$ cases. In each case, the absence of holes in the
    polygon forces visibility relations that we encode as Boolean
    constraints on links and triple orientations. Independently checkable
    SAT refutations exclude all $74$ cases. The upper-bound proof was
    derived using AI. A separate Lean~4 formalization independently
    verifies the upper bound against an axiomatic interface for simple
    polygons.
\end{abstract}

\section{Introduction}
\seclab{intro}
Place six target points in a simple polygon and choose three of them. Is
there a point in the polygon from which precisely those three targets are
visible? Two points in a polygon \emphi{see} each other when the closed
segment between them, called a \emphi{link}, lies completely in the
polygon (including its boundary). A point in the polygon that sees
precisely a chosen subset of targets is a \emphi{witness} for that
subset. If every subset of the targets has a witness, the polygon
\emphi{shatters} the target set, and the largest size of a shattered set
in a polygon is its visibility VC-dimension.

\paragraph{Known bounds and our result.}
Valtr~\cite{Val98} constructed a simple polygon shattering six targets in
strictly convex position and proved an upper bound of $23$ on the
visibility VC-dimension of any simple polygon. Gilbers and
Klein~\cite{GK14} lowered the general upper bound to $14$, and to $9$ when
the targets are in strictly convex position. We prove that no simple
polygon shatters seven targets. Because every subset of a shattered set
is itself shattered, six is the exact maximum across all simple polygons,
both in general and in strictly convex position (\thmref{main-six}).

\paragraph{Guarding and visibility.}
Kalai and Matou\v{s}ek~\cite{KM97} initiated the study of this parameter
by proving that the visibility VC-dimension of every simple polygon is
bounded by an absolute constant. Because visibility is symmetric---a
guard $q$ sees a point $p$ if and only if $q$ lies in the visibility
region of $p$---the $\varepsilon$-net theorem of Haussler and
Welzl~\cite{HW87} then yields $O((1/\varepsilon)\log(1/\varepsilon))$
guards for any simple polygon in which every point sees at least an
$\varepsilon$-fraction of the polygon's area.

\paragraph{From wedges to a finite proof.}
Let $\mV$ be the set of targets on the boundary of the convex hull of a
shattered set. Gilbers and Klein~\cite{GK14} showed that, among the
witnesses seeing all of $\mV$, visibility of each target strictly inside
the hull is governed by membership in an open wedge. Because six wedges
in the plane cannot realize all $2^6 = 64$ membership patterns, at most
five targets can lie strictly inside the hull. We extend this wedge-test
argument to include certain ``certified'' hull vertices and prove that
all certified vertices form a single contiguous block along the hull
boundary. Considering the witness that sees \emph{only} the remaining,
uncertified hull vertices then assigns each of the other targets to a fan
triangle over a hull edge, partitioning all seven-target configurations
into $74$ hull cases (\secref{hull-certified}).

Within each hull case, we represent the triple orientations of targets
and selected witnesses by Boolean variables, in the spirit of the
empty-hexagon proof of Heule and Scheucher~\cite{HS24}, together with
Boolean variables for pairwise visibility links. Because the polygon has
no holes, two polygonal paths of links whose internal vertices lie on
opposite sides of a segment enclose that segment and force it to be a
link as well. Any orientation assignment under which forced links enclose
a sightline that a witness is required to miss is therefore impossible.
Encoding these geometric implications produces $74$ propositional
formulas, each accompanied by an independently checkable proof of
unsatisfiability (\secref{certificates}).

\subsection{Seven targets in convex position}
\seclab{convex-warmup}

As a warm-up, suppose a simple polygon $P$ shatters the vertex set
$\mS=\{v_0,\ldots,v_6\}$ of a strictly convex heptagon $\C=\CHX{\mS}$,
listed in counterclockwise order. Fix one witness $w_\mT$ for each subset
$\mT\subseteq\mS$. By a small perturbation of the polygon and points
(\lemref{general-position}), we may assume that all targets and witnesses
are distinct and no three are collinear.

Suppose all seven sides of $\C$ are links in $P$. We show that no witness
can see precisely $\{v_0,v_2,v_4\}$, contradicting shattering. Because
$P$ has no holes, the seven boundary links enclose $\C$, so $\C\subseteq
P$ (\lemref{enclosure} (i)). By convexity of $\C$, any point in $\C$ sees
all seven targets, so every witness that misses at least one target must
lie in $P \setminus \C$.

Now take a witness $w\in P\setminus\C$ that sees at least two targets.
Since $w\notin\C$, the rays from $w$ to $\C$ span an angle smaller than
$\pi$. Let $b$ and $c$ be the first and last targets seen by $w$ in
counterclockwise angular order around $w$. We claim that $w$ sees every
target $p$ in the closed wedge bounded by the rays $wb$ and $wc$
(\figref{warmup-hull}).

\begin{figure}[htb]
    \centering
    \includegraphics[width=.49\textwidth]{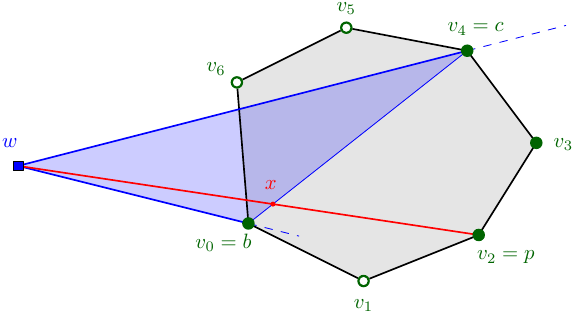}
    \caption{Warm-up on a convex heptagon whose sides all lie in $P$.
        Blue shading marks $\triangle(w,b,c)$, and dashed blue rays
        extend $wb$ and $wc$ to bound the visibility wedge from $w$. The
        ray $wp$ meets the chord $bc$ at $x$: $\triangle(w,b,c)$ covers
        $wx$, while $\C$ covers $xp$. Filled vertices ($v_0$ and
        $v_2,v_3,v_4$) lie in the wedge and are seen by $w$, forming two
        contiguous runs along $\partial\C$.}
    \figlab{warmup-hull}
\end{figure}

Indeed, since $bc\subseteq\C\subseteq P$, all three sides of
$\triangle(w,b,c)$ are links, so $\triangle(w,b,c)\subseteq P$
(\lemref{enclosure} (i)). The ray $wp$ crosses the chord $bc$ at a point
$x$. If $p\in wx$, then $wp\subseteq\triangle(w,b,c)\subseteq P$;
otherwise $wx\subseteq\triangle(w,b,c)\subseteq P$ and
$xp\subseteq\C\subseteq P$, so again $wp\subseteq P$ and $w$ sees $p$.

Slightly widen the wedge around $wb$ and $wc$ without capturing any
additional target. Because $w\notin\C$ and $\C$ is convex, each of the
two boundary rays of the widened wedge crosses $\partial\C$ at most
twice, so a counterclockwise tour of $\partial\C$ enters and leaves the
wedge at most four times in total. Seeing precisely $\{v_0,v_2,v_4\}$,
however, alternates between seen and unseen vertices six times around the
heptagon (once on each side of $v_0$, $v_2$, and $v_4$), a contradiction.
Thus any shattered convex heptagon must have at least one hull edge that
leaves $P$; once a hull edge is blocked by $\partial P$, a witness can
see both of its endpoints without the triangle they span lying inside
$P$.

\subsection{Overview of the proof}
\seclab{overview}

Now suppose $P$ shatters an arbitrary set $\mS$ of seven targets. Fix a
witness family $\W = \Set{w_\mT}{\mT \subseteq \mS}$, one witness per
subset, and apply \lemref{general-position} so that all $7+2^7=135$
points of $\mS\cup\W$ are distinct and no three are collinear. We first
use the visibility patterns required of $\W$ to classify the convex hull
of $\mS$ into $74$ cases, and then refute each case via propositional
constraints on links and triple orientations.

\paragraph{Step 1. Enclosure forces links.}
If two polygonal paths of links join $s$ to $t$ and their internal
vertices lie on opposite open sides of the line $st$, then the two paths
enclose the segment $st$; since $P$ has no holes, $st$ must itself be a
link (\lemref{enclosure} (ii)). Shattering prescribes both positive links
(from $w_\mT$ to every target in $\mT$) and forbidden links (from $w_\mT$
to every target in $\mS\setminus\mT$). Any placement of targets and
witnesses in which forced links enclose a forbidden link is therefore
impossible. \figref{forced-sightline} illustrates the basic mechanism:
two witnesses seeing targets $g$ and $b$ from opposite sides of the line
$gb$ force the link $gb$, after which any witness $w$ seeing $g$ and $b$
must also see every target $t\in\inte\triangle(w,g,b)$.

\begin{figure}[htb]
    \centering
    \includegraphics[width=.74\textwidth]{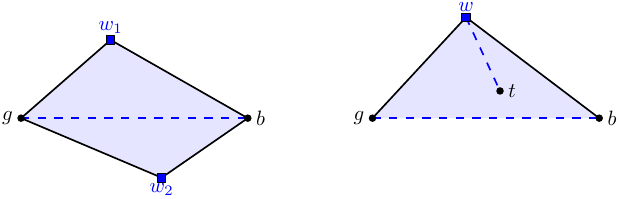}
    \caption{Left: two two-link paths on opposite sides of $gb$ enclose
        and force the dashed link $gb$. Right: once $gb$ is a link, the
        links $wg$, $wb$, and $gb$ enclose $\triangle(w,g,b)$ and force
        the dashed link $wt$.}
    \figlab{forced-sightline}
\end{figure}

\paragraph{Step 2. Counting wedge tests.}
Let $\mV$ be the vertex set of $\CHX{\mS}$ and let $\Sin=\mS\setminus\mV$
be the set of \emphi{interior targets}. Gilbers and Klein~\cite{GK14}
showed that, among the witnesses seeing all of $\mV$, an interior target
$g\in\Sin$ is seen precisely by those witnesses lying in a suitable open
wedge $W_g$ with apex $g$ (\figref{wedge-test}). Since six open wedges
cannot realize all $64$ membership patterns (\thmref{six-wedges}), at
most five targets belong to $\Sin$.

\begin{figure}[htb]
    \centering
    \includegraphics[width=\textwidth]{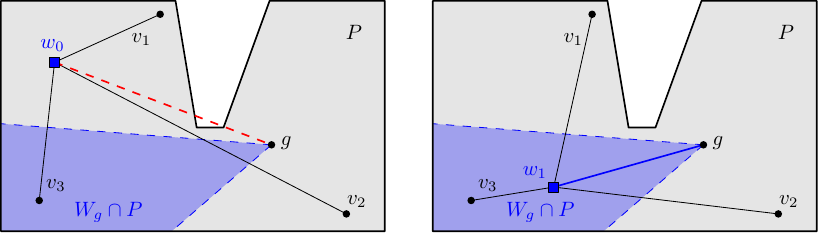}
    \caption{Gilbers and Klein's wedge test~\cite{GK14}. For each target
        $g\in\Sin$ strictly inside $\CHX{\mS}$, an open wedge $W_g$ with
        apex $g$ tests visibility of $g$ among the chosen witnesses that
        see all hull vertices $\mV=\{v_1,v_2,v_3\}$: such a witness $w$
        sees $g$ if and only if $w\in W_g$. Blue shading clips $W_g$ to
        $P$. Left: $w_0\notin W_g$, and the polygon notch blocks $w_0g$.
        Right: $w_1\in W_g$, and $w_1$ sees $g$.}
    \figlab{wedge-test}
\end{figure}

We extend this wedge count to include obstructed hull vertices. Call a
hull vertex $v\in\mV$ \emph{certified} if some witness $w$ sees both hull
neighbors $v_-,v_+$ of $v$, misses $v$, and satisfies
$v\in\inte\triangle(w,v_-,v_+)$ (\figref{certified-vertex}). By the Cone
Lemma of Gilbers and Klein~\cite{GK14} (\lemref{cone}), every witness
seeing both $v_-$ and $v_+$ must then lie in the open cone beyond $v$,
giving $v$ a wedge test alongside the interior targets. We prove that the
set $\mH$ of certified hull vertices forms a single contiguous block on
$\partial\CHX{\mS}$ (\propref{certified-structure}) and that
$\mM = \Sin\cup\mH$ contains at most five targets. Writing
$\mD = \mV\setminus\mH$ for the uncertified hull vertices gives the
partition $\mS = \mD \sqcup \mM$, so $\cardin{\mD} = 7 - \cardin{\mM} \ge
2$.

\begin{figure}[htb]
    \phantom{}\hfill%
    \includegraphics[page=1]{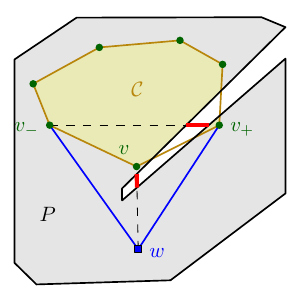}\hfill%
    \includegraphics[page=2]{figs/certified_cone}
    \hfill\phantom{}%
    \caption{A certified hull vertex $v$ of $\C=\CHX{\mS}$. Left: the
        witness $w$ sees the hull neighbors $v_-,v_+$ with
        $v\in\inte\triangle(w,v_-,v_+)$, while a notch of $\partial P$
        blocks $wv$ and $v_-v_+$ (red segments). Right: every witness
        seeing both $v_-$ and $v_+$, such as $w'$, must lie in the shaded
        cone beyond $v$, even when $w'$ also sees $v$.}
    \figlab{certified-vertex}
\end{figure}

\paragraph{Step 3. Exit edges and hull cases.}
Shattering supplies a witness $w_\mD$ that sees precisely $\mD$ and
misses every target in $\mM$. When $\cardin{\mD}\ge3$, the ray from
$w_\mD$ through any missed target $y\in\mM$ leaves $\CHX{\mD}$ across the
interior of a unique \emph{exit edge} $e(y)=u_yv_y$, and every witness
seeing $u_y$ and $v_y$ contains $y$ in the interior of its triangle with
$u_y$ and $v_y$ (\lemref{exit-edge} and \figref{vcdim6-overview}). A hull
case records $\cardin{\mD}$, $\cardin{\mH}$, and (when
$\cardin{\mD}\ge3$) these exit edges; modulo dihedral symmetries of
$\CHX{\mD}$, there are $74$ cases (\secref{case-enumeration}).

\begin{figure}[htb]
    \centering
    \includegraphics[width=0.62\textwidth]{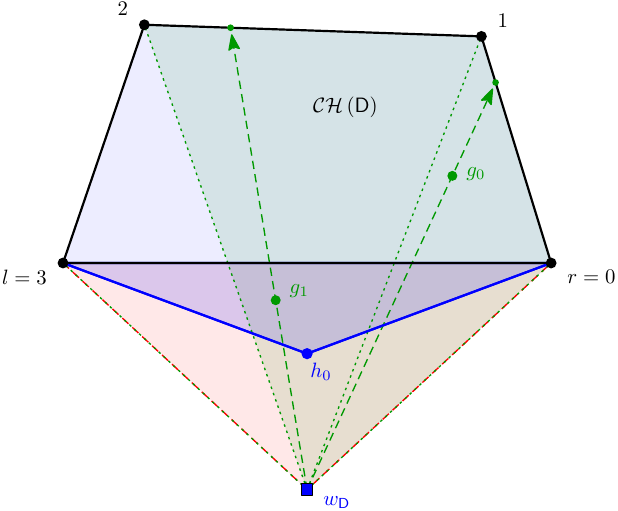}
    \caption{Exit-edge classification for $\mD=\{0,1,2,3\}$ (black
        vertices), one certified hull vertex $h_0$ (blue) between its
        hull neighbors $l=3$ and $r=0$, and two interior targets
        $g_0,g_1$ (green). Dotted green segments from $w_\mD$ to $1$ and
        $2$ divide $\CHX{\mD\cup\{w_\mD\}}$ into three fan triangles. The
        dashed ray from $w_\mD$ through $g_0$ exits $\CHX{\mD}$ across
        $01$, while $h_0$ and the ray through $g_1$ (entering across $30$)
        exit across $12$; thus $e(g_0)=01$ and $e(h_0)=e(g_1)=12$.}
    \figlab{vcdim6-overview}
\end{figure}

\paragraph{Step 4. A Boolean formula excludes each case.}
For each hull case we select a witness subfamily $\W_0\subseteq\W$
containing the $\binom{7}{0}+\binom{7}{1}+\binom{7}{2}=29$ witnesses that
miss at most two targets together with a handful of case-specific
witnesses such as $w_\mD$. Any realizing polygon determines the
orientation sign of every triple in $\mS\cup\W_0$ and the visibility of
every pair. Encoding both as Boolean variables turns orientation
identities, the hull order, the enclosing triangles of Step~3, a
half-plane ordering constraint, and the enclosure rules of Step~1 into
propositional clauses (\secref{constraints}). All $74$ resulting formulas
are unsatisfiable, as certified by independently verified DRAT
refutations.

\paragraph{Checking the upper bound.}
Our supplementary repository (\url{https://github.com/sarielhp/vc_vis_6})
archives the formula generator, a manifest of SHA-256 digests, and a DRAT
refutation for each of the $74$ cases. A replay script checks the case
enumeration, regenerates each formula, independently verifies the
geometric derivation of each saved cut clause, checks the SHA-256 digest,
and validates the refutation using \texttt{drat-trim} of Wetzler
\etal~\cite{WHH14}. This reduces the computational trust base to the
formula generator and the DRAT checker; an independent Lean~4
formalization eliminates the former as well (\secref{certificates}).
\figref{proof-pipeline} summarizes the resulting proof pipeline.

\begin{figure}[htb]
    \centering
    \includegraphics[width=\textwidth]{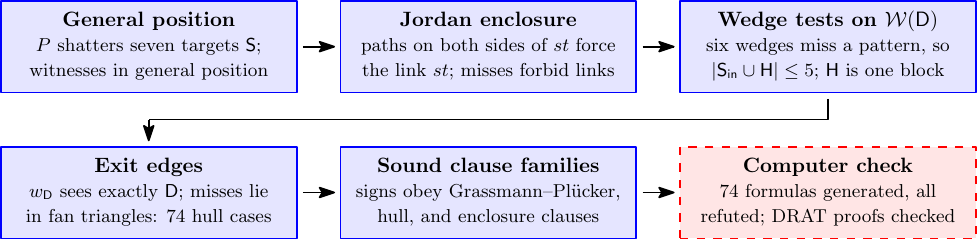}
    \caption{Architecture of the upper-bound proof. Solid boxes summarize
        the geometric reductions and clause derivations: $\Sin$ denotes
        the interior targets, $\mH$ the certified hull vertices, $\mD$
        the remaining hull vertices, and $\W(\mD)$ the witnesses seeing
        all of $\mD$. For each of the $74$ hull cases, the dashed box
        verifies an independent DRAT refutation (\secref{certificates}).}
    \figlab{proof-pipeline}
\end{figure}

\paragraph{AI Statement.}
This proof was found with an experimental search harness on Gemini 4
Carbon. The authors built the search harness, steered the search,
verified the proof, and wrote the manuscript. The computational part of
the proof is certified by proofs that are checked by an independent
program (\secref{certificates}). The upper bound was also formally
verified in Lean~4, assuming an interface for simple polygons
(\secref{certificates}). The AI was also used in the writing of this
manuscript.

\paragraph{Organization.}
\secref{prelim} sets up the general-position reduction, wedge counting,
Jordan enclosure, and wedge extraction. \secref{hull-certified} proves
that the certified hull vertices form a single block, defines exit edges,
and enumerates the $74$ hull cases. \secref{constraints} establishes the
four-cycle, iterated-enclosure, and half-plane visibility lemmas and
proves soundness of all $11$ propositional clause families.
\secref{proof-main} completes the proof of \thmref{main-six} and
describes the DRAT and Lean~4 certificates. \apndref{general-position}
proves the general-position reduction.

\FloatBarrier
\section{Visibility and wedge tests}
\seclab{prelim}
To bound how many targets can be shattered simultaneously, we need a
single witness subfamily on which several targets admit simultaneous
\emph{wedge tests}---open wedges whose membership characterizes which
witnesses in the subfamily see each target. After perturbing the
configuration into general position, we review the wedge-counting bound
and show how any target missed inside a triangle of seen targets admits a
wedge test.

\subsection{Visibility, shattering, and orientation predicates}

For three points $p, q, r \in \R^2$, let
\[
    \orient(p, q, r) = \det(q - p, r - p),
\]
which is positive if and only if $(p, q, r)$ makes a strict left
(counterclockwise) turn. We write $\triangle(p, q, r)$ for the closed
triangle with vertices $p, q, r$, $\CHX{\mX}$ for the convex hull of $\mX
\subset \R^2$, and
\[
    \cone(d_1, d_2) = \Set{\alpha d_1 + \beta d_2}{\alpha, \beta > 0}
\]
for the open cone spanned by two linearly independent vectors
$d_1,d_2\in\R^2$. A line $\ell$ \emphi{separates} points $p$ and $q$ if
they lie in opposite open half-planes bounded by $\ell$.

\begin{definition}
    \deflab{visibility}
    Let $P \subset \R^2$ be a simple polygon (a compact region bounded by
    a simple closed polygonal curve $\partial P$). A closed segment
    $pq\subseteq P$ is a \emphi{link} in $P$, and its endpoints $p,q\in P$
    \emphi{see} each other if and only if $pq$ is a link. The
    \emphi{visibility region} of $p\in P$ is
    \[
        \vis_P(p) = \Set{q\in P}{pq\subseteq P}.
    \]
    We write $\vis(p) = \vis_P(p)$ when $P$ is clear from context.
\end{definition}

\begin{definition}
    \deflab{shattered}
    A finite set $\mS \subset P$ is \emphi{shattered} in $P$ if for every
    subset $\mT \subseteq \mS$ there exists a \emphi{witness} $w_\mT \in P$
    satisfying $\vis_P(w_\mT) \cap \mS = \mT$. Fixing one witness $w_\mT$ for
    each $\mT \subseteq \mS$ defines a \emphi{witness family} $\W =
    \Set{w_\mT}{\mT \subseteq \mS}$. For disjoint subsets $\mF, \mN \subseteq \mS$,
    let
    \begin{align*}
        \witnessfamily{\mF}{\mN} &= \Set{w \in \W}{\mF \subseteq \vis_P(w) \text{ and } \mN
        \cap \vis_P(w) = \emptyset},\\
        \W(\mF) &= \witnessfamily{\mF}{\emptyset}
    \end{align*}
    denote the subfamily of witnesses in $\W$ that see every target in
    $\mF$ and miss every target in $\mN$, and the subfamily of witnesses
    that see all of $\mF$, respectively.
\end{definition}

\begin{definition}
    \deflab{vcdim}
    The \emphi{visibility VC-dimension} of a simple polygon $P$ is
    \[
        \vcdim(P) = \max\Set{\cardin{\mS}}{\mS \subset P \text{ is
        shattered in } P}.
    \]
    A finite set $\mS \subset \R^2$ is in \emphi{strictly convex
    position} if $s \notin \CHX{\mS \setminus \{s\}}$ for every $s \in
    \mS$, and the \emphi{convex-position visibility VC-dimension}
    $\vcdimconv(P)$ is the maximum cardinality $\cardin{\mS}$ of a
    shattered set $\mS \subset P$ in strictly convex position.
\end{definition}

\begin{observation}
    \obslab{monotonicity}
    If $\mS \subset P$ is shattered in $P$ with witness family $(w_\mT)_{\mT
    \subseteq \mS}$, then every subset $\mS_0 \subseteq \mS$ is also shattered
    in $P$, since $\vis_P(w_{\mT_0}) \cap \mS_0 = \mT_0 \cap \mS_0 = \mT_0$ for
    every $\mT_0 \subseteq \mS_0$. In particular, if $\vcdim(P) \ge 7$, then
    $P$ contains a shattered set of cardinality~$7$.
\end{observation}

\begin{definition}
    \deflab{test}
    Let $g \in \mS$ and $\W' \subseteq \W$. An open set $R \subseteq \R^2$
    is a \emphi{test} for $g$ on $\W'$ if $\partial R \cap \W' =
    \emptyset$ and
    \[
        w \text{ sees } g \iff w \in R \qquad \text{for every } w \in \W'.
    \]
    The test $R$ is a \emphi{wedge test} if $R$ is a translate of
    $\cone(d_1,d_2)$ for some linearly independent $d_1,d_2\in\R^2$.
\end{definition}

\subsection{General-position reduction}

Before analyzing visibility constraints, we perturb any shattered
configuration into the interior of a slightly enlarged simple polygon so
that no three points are collinear and every positive visibility link has
open-disk slack at its witness endpoint.

\begin{lemma}
    \lemlab{general-position}
    If a simple polygon $P \subset \R^2$ shatters a finite set $\mS \subset
    P$, then there exist a simple polygon $P' \supset P$, a shattered set
    $\mS' \subset \inte P'$ with $\cardin{\mS'} = \cardin{\mS}$ (in strictly
    convex position whenever $\mS$ is) with a bijection $s \mapsto s'$ from
    $\mS$ to $\mS'$, and witnesses $\W' = \Set{w'_\mT}{\mT \subseteq \mS} \subset
    \inte P'$ realizing $\vis_{P'}(w'_\mT) \cap \mS' = \Set{s'}{s \in \mT}$
    with the following properties:
    \begin{conditions}{G}
    \item\condlab{general-position} The set $\mS'\cup\W'$ consists of
        $\cardin{\mS'}+2^{\cardin{\mS'}}$ distinct points, with no three collinear.
    \item\condlab{witness-disks} For every witness $w\in\W'$ there is an open disk
        $D_w\subset\inte P'$ centered at $w$ such that every point in $D_w$
        sees every target in $\vis_{P'}(w)\cap \mS'$.
    \end{conditions}
\end{lemma}

\begin{proofidea}
    Inflate $\partial P$ outward slightly so that existing links gain
    positive clearance from $\partial P'$ while every blocked witness--target
    segment still leaves $P'$, and then perturb the points of $\mS$ and
    $\W$ one by one within that clearance off all previously spanned
    lines. The full proof appears in \apndref{general-position}.
\end{proofidea}

Throughout the remainder of the paper, whenever $\mS \subset P$ is a
shattered set, we apply \lemref{general-position} first and write $(P,
\mS, \W)$ directly for the resulting general-position configuration
satisfying \condref[]{general-position}--\condref[]{witness-disks}.

\subsection{Counting wedge tests}
\seclab{counting}

Because $\witnessfamily{\mF'}{\mN'} \subseteq \witnessfamily{\mF}{\mN}$
whenever $\mF \subseteq \mF'$ and $\mN \subseteq \mN'$, any test for a
target on $\witnessfamily{\mF}{\mN}$ restricts to a test on
$\witnessfamily{\mF'}{\mN'}$. Simultaneous tests on a single witness
subfamily $\witnessfamily{\mF}{\mN}$ force the arrangement of test regions
to realize every subset pattern:

\begin{lemma}
    \lemlab{counting}
    Let $\mF, \mN \subseteq \mS$ be disjoint, and let $g_1, \dots, g_\nu \in \mS
    \setminus (\mF \cup \mN)$ be distinct. If $R_\mu$ is a test for $g_\mu$ on
    $\witnessfamily{\mF}{\mN}$ for each $\mu \in \{1, \dots, \nu\}$, then for every $\mK
    \subseteq \{1, \dots, \nu\}$ there exists a point $p$ outside
    $\bigcup_{\mu=1}^\nu \partial R_\mu$ with $\Set{\mu}{p \in R_\mu}=\mK$.
\end{lemma}

\begin{proof}
    Take $p = w_\mT \in \witnessfamily{\mF}{\mN}$ for $\mT = \mF \cup
    \Set{g_\mu}{\mu \in \mK}$. By \defref{test}, $w_\mT$ avoids every
    boundary $\partial R_\mu$ and belongs to $R_\mu$ if and only if $w_\mT$
    sees $g_\mu$, which holds precisely when $\mu \in \mK$.
\end{proof}

Isler \etal~\cite{Isl+04} bounded the number of visibility patterns of
rays in the exterior of a convex hull; Gilbers and Klein~\cite{GK14}
reformulated their bound for arrangements of open wedges.

\begin{theorem}
    \thmlab{six-wedges}
    For any six open wedges $R_1, \dots, R_6 \subset \R^2$, there is a
    subset $\mK \subseteq \{1,\dots,6\}$ for which no point $p$ outside
    $\bigcup_{\mu=1}^6 \partial R_\mu$ satisfies $\Set{\mu}{p\in R_\mu}=\mK$.
\end{theorem}

\noindent We give a short self-contained proof via Euler's formula.

\begin{proof}
    Pick one witness point outside $\bigcup_{\mu=1}^6 \partial R_\mu$ for
    each realized membership pattern. Because there are at most $2^6$
    such points and each lies strictly inside or strictly outside every
    wedge, we may slightly perturb the six apices and twelve boundary rays
    without losing any realized pattern so that the six apices are
    distinct, the twelve boundary rays have distinct directions and
    supporting-line slopes, no three boundary lines are concurrent, and no
    apex lies on a boundary line of another wedge.

    \emph{Count the cells.} Compactify $\R^2$ to a closed disk by adding a
    circle $S^1$ of directions at infinity. Let $c_{\mu\nu}$ be the number
    of crossings between $\partial R_\mu$ and $\partial R_\nu$, and put
    $C = \sum_{1\le\mu<\nu\le 6} c_{\mu\nu}$. The arrangement on the disk
    has $V = 6 + 12 + C = 18 + C$ vertices ($6$ apices, $12$ ray endpoints
    on $S^1$, and $C$ ray crossings) and $E = 12 + (12 + 2C) = 24 + 2C$
    edges ($12$ arcs along $S^1$ and $12 + 2C$ ray segments). Every apex
    and crossing is connected along boundary rays to $S^1$, so this planar
    graph is connected. By Euler's formula on the disk, $V - E + F = 1$,
    giving $F = 7 + C$ two-dimensional cells.

    Suppose two of the twelve arcs along $S^1$ border the same cell. A
    simple curve through that cell joining the interiors of those two
    arcs cuts the disk into two parts, each containing at least one ray
    endpoint on $S^1$. Because the two boundary rays of any wedge meet at
    its apex, both rays lie on the same side of the cut. The six wedges
    therefore split into two nonempty groups with no crossings between
    groups. At most $\binom{5}{2} = 10$ pairs of wedges lie in the same
    group, and each pair of two-ray boundaries crosses at most
    $c_{\mu\nu} \le 4$ times, so $C \le 40$ and $F \le 47 < 64$. We may
    therefore assume that all twelve arcs along $S^1$ border distinct
    cells.

    \emph{Count the distinct patterns.} Let $I_\mu \subset S^1$ be the
    open arc of directions of $R_\mu$, which has angular length less than
    $\pi$. We claim that if $c_{\mu\nu} = 4$, then $I_\mu \cap I_\nu =
    \emptyset$. Let $d \in S^1$ be the direction from the apex of $R_\nu$
    to the apex of $R_\mu$. A ray from the apex of $R_\nu$ crosses a
    boundary ray of $R_\mu$ with direction $\theta$ if and only if its
    direction lies in the open shorter arc of $S^1$ from $d$ to $\theta$.
    For both boundary rays of $R_\nu$ to cross both boundary rays of
    $R_\mu$, both boundary directions of $I_\nu$ must lie in the
    intersection of the two swept arcs from $d$ to the endpoints of
    $I_\mu$. Hence both endpoints of $I_\mu$ lie on the same side of $d$,
    at angular distances $0 < \theta_1 < \theta_2 < \pi$ from $d$: the
    common swept arc is $(0, \theta_1)$ while $I_\mu$ is $(\theta_1,
    \theta_2)$. Since both endpoints of $I_\nu$ lie in $(0, \theta_1)$
    and $I_\nu$ has length less than $\pi$, the entire arc $I_\nu$ lies in
    $(0, \theta_1)$ and is disjoint from $I_\mu$.

    Let $e$ and $\kappa$ be the numbers of edges and connected components
    of the intersection graph of the six open arcs $I_1, \dots, I_6$.
    Each intersecting pair has $c_{\mu\nu} \le 3$, so $C \le 4\binom{6}{2}
    - e = 60 - e$ and $F \le 67 - e$. If $\bigcup_{\mu=1}^6 I_\mu = S^1$,
    choose a minimal subcover; it uses at least three arcs since each has
    length less than $\pi$. Pick a point private to each arc of the
    minimal subcover. The gap between two consecutive private points can
    be covered only by their two arcs (any third subcover arc reaching the
    gap from its own private point would cover one of the two private
    points), so adjacent arcs in the subcover overlap, forming a cycle.
    Because $S^1$ is connected, the six-vertex intersection graph is
    connected and contains a cycle, giving $e \ge 6$ and $F \le 61$.
    Otherwise $S^1 \setminus \bigcup_{\mu=1}^6 I_\mu$ is nonempty; because
    all twelve arc endpoints are distinct, it consists of $\kappa$
    disjoint closed gaps. Each gap contains one of the twelve $S^1$-arcs,
    and points sufficiently far out in an interior direction of a gap miss
    all six wedges. Thus $\kappa$ distinct cells realize the empty
    pattern $\emptyset$, leaving at most
    \[
        F - (\kappa - 1) \le 68 - (e + \kappa) \le 62
    \]
    distinct patterns, since every six-vertex graph satisfies $e + \kappa
    \ge 6$. In every case, fewer than $64$ patterns occur.
\end{proof}

\subsection{Enclosure and the Cone Lemma}
\seclab{enclosure-cone}

For non-collinear points $g, b, c \in \R^2$, define the \emphi{opposite cone}
\[
    \oppcone{g}{b}{c} = g + \cone(g - b, g - c).
\]
Equivalently, $x \in \oppcone{g}{b}{c}$ if and only if $g \in \inte\triangle(x,
b, c)$.

Because a simple polygon $P$ has no holes, any simple closed polygonal
curve in $P$ keeps its interior inside $P$; we refer to the two assertions
below as \emph{Jordan enclosure}.

\begin{lemma}
    \lemlab{enclosure}
    \leavevmode
    \begin{compactenumi}
    \item If a simple closed polygonal curve $\Gamma$ lies in $P$, then
        the bounded region enclosed by $\Gamma$ lies in $P$. In
        particular, if $p,q,r\in P$ pairwise see each other, then
        $\triangle(p,q,r)\subseteq P$.
    \item Let $\pi_1,\pi_2$ be polygonal paths in $P$ joining distinct
        points $s,t$, each with at least two edges. If the internal
        vertices of $\pi_1$ and of $\pi_2$ lie in opposite open
        half-planes bounded by the line $st$, then $s$ sees $t$.
    \end{compactenumi}
\end{lemma}

\begin{proof}
    (i)~The connected unbounded exterior $\R^2\setminus P$ is disjoint
    from $\Gamma$ and therefore lies in the unbounded component of
    $\R^2\setminus\Gamma$.
    (ii)~Shortcut any self-intersections in $\pi_1$ and $\pi_2$ to obtain
    simple polygonal paths in $P$ from $s$ to $t$. By convexity of open
    half-planes, apart from the endpoints $s,t$, each shortcut path stays
    in the open half-plane containing the internal vertices of the
    original path. Their union is therefore a simple closed polygonal
    curve in $P$ meeting the line $st$ only at $s$ and $t$. Its enclosed
    bounded region contains the open segment $st\setminus\{s,t\}$, so~(i)
    gives $st\subseteq P$.
\end{proof}

When \lemref{enclosure} (ii) applies, we say that $\pi_1$ and $\pi_2$
\emphi{enclose} the segment $st$.

The next result, the \emph{Cone Lemma} of Gilbers and Klein~\cite{GK14},
shows that once a witness sees $b$ and $c$ while missing a point $g$
inside $\triangle(w,b,c)$, every other witness seeing $b$ and $c$ is
confined to the opposite cone $\oppcone{g}{b}{c}$.

\begin{lemma}
    \lemlab{cone}
    Let $g,\allowbreak b,\allowbreak c,\allowbreak w \in P$ satisfy $b, c \in \vis(w)$, $g \notin
    \vis(w)$, and $g \in \inte\triangle(w, b, c)$. If $w' \in P$ sees $b$
    and $c$ and lies on none of the lines $gb, gc, wg$, then $w' \in \oppcone{g}{b}{c}$.
\end{lemma}

\begin{figure}[htb]
    \centering
    \includegraphics[width=.43\textwidth]{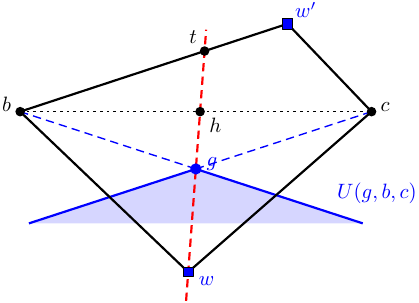}
    \caption{Proof of \lemref{cone}. If $w'$ lay outside the blue cone
        $\oppcone{g}{b}{c}$ and saw both $b$ and $c$, the dashed red line
        $\ell = wg$ would cross $bc$ at $h$ and $bw'$ at $t$ with $g$
        between $w$ and $t$; the paths $w-b-t$ and $w-c-w'-t$ would then
        enclose $wt$ and force $w$ to see $g$.}
    \figlab{cone}
\end{figure}

\begin{proof}
    Let $\ell$ be the line $wg$, which separates $b$ from $c$ and meets
    $bc$ at $h$ with $g$ strictly between $w$ and $h$ (\figref{cone}).
    Since $w' \notin \ell$, the two-link path $b - w' - c$ crosses $\ell$
    once, say on $bw'$ at $t \neq g$. If $g$ lay between $w$ and $t$,
    then the paths $w - b - t$ and $w - c - w' - t$ would enclose $wt$,
    forcing $w$ to see $t$ and hence $g$, a contradiction. Thus $g$ lies
    strictly between $t$ and $h$; since $t$ and $h$ lie in the relative
    interiors of $bw'$ and $bc$, this places $g \in \inte\triangle(w', b,
    c)$, i.e., $w' \in \oppcone{g}{b}{c}$.
\end{proof}

\subsection{Wedge extraction}
\seclab{wedge-extraction}

Whenever every witness in a subfamily $\W' \subseteq \W(\{b,c\})$ that
sees $g$ is confined to a cone $U$ whose rays through $g$ separate $b$
from $c$, Gilbers and Klein~\cite{GK14} showed that $g$ admits a wedge
test on $\W'$. Every wedge test in \secref{hull-certified} arises from
this lemma.

\begin{lemma}
    \lemlab{wedge-extraction}
    Let $g,b,c\in \mS$ be distinct, and let $\W'\subseteq\W(\{b,c\})$ be
    any subfamily of witnesses seeing both $b$ and $c$. Let $U$ be an open
    wedge with apex $g$ such that every line through $g$ and a point of
    $U$ separates $b$ from $c$. If every witness in $\W'$ seeing $g$ lies
    in $U$, then $g$ has a wedge test $W \subseteq U$ with apex $g$ on
    $\W'$.
\end{lemma}

\begin{proof}
    Let $\W_{+}$ be the set of witnesses in $\W'$ that see $g$. We span
    $W$ just outside the extreme rays from $g$ through $\W_{+}$ so that
    any witness of $\W'$ falling inside $W$ is enclosed between link paths
    through points near those extreme witnesses.

    If $\W_{+}=\emptyset$, the finite set $\W'$ determines finitely many
    ray directions from $g$ (as $g\notin\W'$ since $\mS$ and $\W$ are
    disjoint by \condref[]{general-position}), so we may choose a narrow open
    wedge $W\subseteq U$ with apex $g$ whose closure misses $\W'$.

    Suppose $\W_{+}\ne\emptyset$ (\figref{wedge-extraction}). Take the
    first and last rays from $g$ through $\W_{+}$ in angular order within
    $U$. Choose points $p_1,p_2 \in U$ just outside these two extreme
    rays, inside the open disks given by \condref[]{witness-disks} around
    the corresponding extreme witnesses (if $\cardin{\W_{+}}=1$, pick
    $p_1$ and $p_2$ in that single witness's disk on opposite sides of its
    ray from $g$). Because every point in a witness's disk of
    \condref[]{witness-disks} sees all targets seen by that witness, both
    $p_1$ and $p_2$ see $g,b,c$. Perturbing $p_1,p_2$ slightly if needed,
    ensure that neither equals a target and neither ray $gp_1, gp_2$
    passes through a point of $\W'$. The open wedge $W \subseteq U$
    between $gp_1$ and $gp_2$ then contains $\W_{+}$ and satisfies
    $\partial W\cap\W'=\emptyset$.

    Conversely, fix $w\in\W'\cap W$. The line $gw$ separates $p_1$ from
    $p_2$ and, by the hypothesis on $U$, separates $b$ from $c$. Relabel
    $\{b,c\}$ if needed so that $p_1,b$ lie on one open side of $gw$ and
    $p_2,c$ lie on the other. Since $p_1$ sees $g,b$, $p_2$ sees $g,c$,
    and $w\in\W'$ sees $b,c$, the link paths $g-p_1-b-w$ and $g-p_2-c-w$
    lie in $P$ on opposite sides of $gw$. By \lemref{enclosure} (ii) they
    enclose $gw$, so $w\in\W_{+}$ and hence $\W'\cap W=\W_{+}$.
\end{proof}

\begin{figure}[htb]
    \centering
    \includegraphics[width=.5\textwidth]{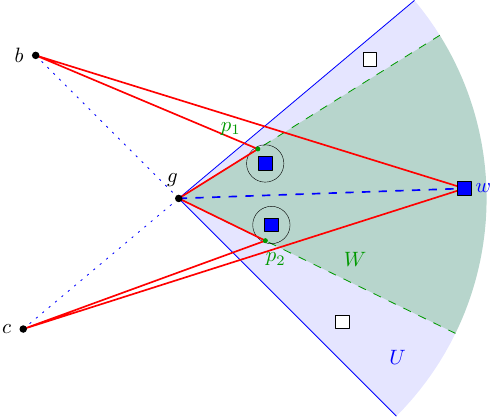}
    \caption{\lemref{wedge-extraction} with $U=\oppcone{g}{b}{c}$. Filled
        squares in $\W'$ see $g$; hollow squares miss $g$. Points
        $p_1,p_2$, chosen inside the disks of \condref[]{witness-disks}
        around the extreme witnesses seeing $g$, bound the green wedge
        $W$. For any $w\in\W'\cap W$, the red link paths $g-p_1-b-w$ and
        $g-p_2-c-w$ enclose and force the dashed link $gw$.}
    \figlab{wedge-extraction}
\end{figure}

\begin{figure}[htb]
    \centering
    \includegraphics[width=.48\textwidth]{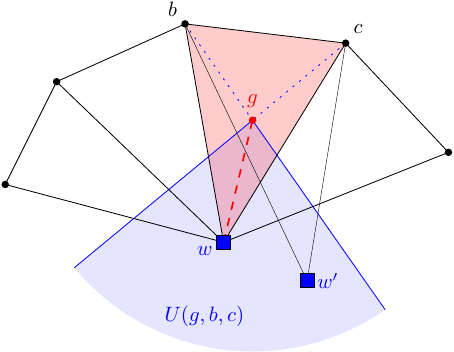}
    \caption{\corref{fan}. Triangles $\triangle(w,b,c)$ with $b,c\in\mF$
        (black targets) fan from $w$ to cover $\CHX{\mF\cup\{w\}}$. Since
        $w$ misses $g \in \inte\triangle(w,b,c)$ (red), \lemref{cone}
        places every witness seeing $b$ and $c$, such as $w$ and $w'$, in
        the blue cone $\oppcone{g}{b}{c}$.}
    \figlab{fan-cover}
\end{figure}

\begin{corollary}
    \corlab{fan}
    Let $w \in \W$, let $\mF \subseteq \vis(w) \cap \mS$, and let $g \in \mS
    \setminus \vis(w)$. If $g \in \CHX{\mF \cup \{w\}}$, then $g$ has a
    wedge test on $\W(\mF)$.
\end{corollary}

\begin{proof}
    Triangles $\triangle(w,b,c)$ with $b,c\in \mF$ cover
    $\CHX{\mF\cup\{w\}}$ (\figref{fan-cover}), so
    \condref[]{general-position} gives $g\in \inte\triangle(w, b, c)$ for
    some $b, c \in \mF$. By \lemref{cone} and
    \condref[]{general-position}, every witness in $\W(\{b, c\})$ seeing
    $g$ lies in $\oppcone{g}{b}{c}$, and every line through $g$ and a
    point of $\oppcone{g}{b}{c}$ separates $b$ from $c$. Applying
    \lemref{wedge-extraction} with $U=\oppcone{g}{b}{c}$ gives a wedge
    test for $g$ on $\W(\{b, c\}) \supseteq \W(\mF)$.
\end{proof}

For the propositional encoding in \secref{constraints}, we also record the
direct orientation consequence of \lemref{wedge-extraction}: no witness
that sees $u,v$ and misses $y$ can fall in the open cone at $y$ spanned by
two witnesses that see $u,v,y$.

\begin{lemma}
    \lemlab{wedge-sep}
    Let $y \in \mS$, let $u, v \in \mS \setminus \{y\}$ be distinct, and
    suppose $y \in \inte\triangle(w, u, v)$ for every witness $w \in
    \W(\{u, v\})$ that sees $y$. Then for any two distinct witnesses
    $w_1, w_2 \in \W(\{u, v, y\})$ and any witness $w_0 \in \witnessfamily{\{u, v\}}{\{y\}}$, we have $w_0 \notin y + \cone(w_1 - y, w_2 - y)$.
\end{lemma}

\begin{proof}
    By \condref[]{general-position}, $w_1 - y$ and $w_2 - y$ are linearly independent.
    By \lemref{wedge-extraction} with $U = \oppcone{y}{u}{v}$, there is an open
    wedge $W \subseteq \oppcone{y}{u}{v}$ with apex $y$ containing $\W(\{u, v,
    y\})$ and disjoint from $\witnessfamily{\{u, v\}}{\{y\}}$. By convexity of $W$,
    $y + \cone(w_1 - y, w_2 - y) \subseteq W$, so $w_0 \notin y + \cone(w_1
    - y, w_2 - y)$.
\end{proof}
\section{The certified hull block and the exit-edge fan}
\seclab{hull-certified}
We now identify the hull vertices whose visibility can be tested
simultaneously with that of the interior targets. The structural results
of \secref{hidden-front}--\secref{exit-edges} hold for any shattered set
$\mS$ of size $\cardin{\mS}\ge 3$; we specialize to $\cardin{\mS}=7$ in
\secref{case-enumeration}.

Put
\[
    \C = \CHX{\mS}, \qquad \mV = \partial\C \cap \mS, \qquad \Sin = \mS
    \setminus \mV.
\]
By \condref[]{general-position}, $\mV$ is the vertex set of $\C$ (with
$\cardin{\mV}\ge 3$ whenever $\cardin{\mS}\ge 3$), and every target of
$\Sin$ lies strictly in $\inte\C$. The witness $w_\mV \in \W$ sees
precisely the hull vertices $\mV$ and misses all of $\Sin$.

\subsection{Hidden targets and front vertices}
\seclab{hidden-front}

Gilbers and Klein~\cite{GK14} used a triangle fan from $w_\mV$ to obtain
simultaneous wedge tests for all interior targets on $\W(\mV)$, proving
$\cardin{\Sin}\le 5$. The same argument extends to a subset $\mT \subseteq
\mV$ of hull vertices whenever each $t \in \mT$ lies inside a triangle
spanned by some witness missing $t$ and two hull vertices outside $\mT$:

\begin{lemma}
    \lemlab{hidden-targets}
    Let $\mT \subseteq \mV$. Suppose that for every $t \in \mT$ there exist
    $b_t, c_t \in \mV \setminus \mT$ and a witness $w^\circ_t \in \W$ such that
    $w^\circ_t$ sees $b_t, c_t$, misses $t$, and $t \in \inte\triangle(w^\circ_t,
    b_t, c_t)$. Then $\cardin{\Sin} + \cardin{\mT} \le 5$. In
    particular, setting $\mT = \emptyset$ gives $\cardin{\Sin}
    \le 5$.
\end{lemma}

\begin{proof}
    Put $\mF=\mV\setminus \mT$. We construct wedge tests for all targets in
    $\Sin \cup \mT$ on the common witness subfamily $\W(\mF)$. For each
    $t\in \mT$, the witness $w^\circ_t$ sees $b_t,c_t$ and misses
    $t\in\inte\triangle(w^\circ_t,b_t,c_t)$, so \corref{fan} gives a wedge
    test for $t$ on $\W(\{b_t,c_t\}) \supseteq \W(\mF)$.

    For the interior targets $\Sin$, consider the witness $w_\mV$, which
    sees all of $\mV$ and misses all of $\Sin$. For each $t\in \mT$,
    $w_\mV$ sees $b_t,c_t \in \mF$, so applying \lemref{cone} to
    $w^\circ_t$ yields $t \in \inte\triangle(w_\mV,b_t,c_t)$ (immediately
    if $w_\mV=w^\circ_t$). Hence $\CHX{\mF\cup\{w_\mV\}}$ contains
    $\mF\cup \mT=\mV$ and therefore contains $\C = \CHX{\mV}$. Every
    $g\in\Sin$ thus lies in $\CHX{\mF\cup\{w_\mV\}}$, so \corref{fan}
    applied to $w_\mV$ and $\mF$ gives a wedge test for $g$ on $\W(\mF)$.
    If $\cardin{\Sin} + \cardin{\mT} \ge 6$, choosing six targets in
    $\Sin\cup\mT$ would realize all $64$ membership patterns of six open
    wedges by \lemref{counting}, contradicting \thmref{six-wedges}.
\end{proof}

To describe which hull vertices satisfy the hypothesis of
\lemref{hidden-targets}, we record the portion of a convex polygon's
boundary that faces an external viewpoint (\figref{front-vertices}).

\begin{definition}
    \deflab{front-vertex}
    Let $\mV' \subset \R^2$ be a finite set in strictly convex position
    with $\cardin{\mV'} \ge 3$, and let $p \in \R^2$ lie on no line
    through two points of $\mV'$. An edge $bc$ of $\CHX{\mV'}$
    \emphi{faces} $p$ if the line $bc$ separates $p$ from
    $\mV'\setminus\{b,c\}$. A vertex $v\in \mV'$ is a \emphi{front vertex}
    of $\mV'$ from $p$ if both edges of $\CHX{\mV'}$ incident to $v$ face
    $p$, and a \emphi{tangent vertex} if exactly one incident edge faces
    $p$. When $p \notin \CHX{\mV'}$, the edges of $\CHX{\mV'}$ facing $p$
    form a single contiguous polygonal chain along $\partial\CHX{\mV'}$,
    called the \emphi{near chain} $N_p$ of $\mV'$ from $p$.
\end{definition}

If $p \in \CHX{\mV'}$, no edge of $\CHX{\mV'}$ faces $p$. If $p \notin
\CHX{\mV'}$, the near chain $N_p$ runs between two distinct tangent
vertices, its internal vertices are precisely the front vertices from
$p$, and all vertices along $N_p$ appear in strictly monotone angular
order around $p$ (\figref{front-vertices}). Moreover, a vertex $v \in
\mV'$ with hull neighbors $v_-, v_+$ in $\mV'$ is a front vertex from $p$
if and only if $p \in \oppcone{v}{v_-}{v_+}$, i.e., $v \in
\inte\triangle(p, v_-, v_+)$.

\begin{figure}[htb]
    \centering
    \includegraphics[width=0.68\textwidth]{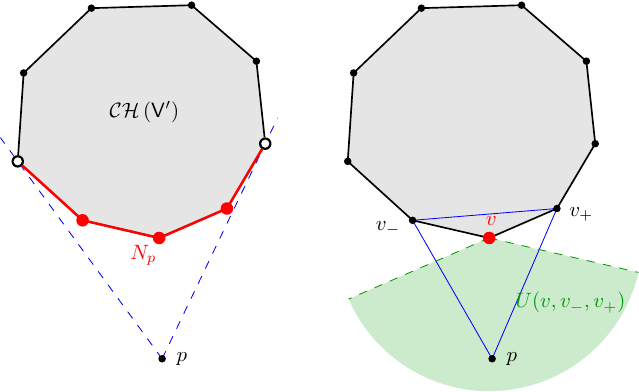}
    \caption{Left: the edges of $\CHX{\mV'}$ facing $p$ form the red near
        chain $N_p$, with front vertices as internal vertices (filled red)
        and tangent vertices as endpoints (hollow circles). Right: a
        vertex $v$ with hull neighbors $v_-,v_+$ is a front vertex from
        $p$ if and only if $v\in\inte\triangle(p,v_-,v_+)$ (blue triangle),
        equivalently $p\in\oppcone{v}{v_-}{v_+}$ (green shaded cone
        between the dashed green extensions).}
    \figlab{front-vertices}
\end{figure}

\begin{lemma}
    \lemlab{hidden-front}
    Assume $\cardin{\mV} \ge 3$. Let $w \in \W$, and let $f \in \mV$ be a
    front vertex of $\mV$ from $w$ with neighbors $b, c$ in $\mV$. If $w$
    sees $b, c$ and misses $f$, then every witness $w' \in \W(\{b,c\})$ lies
    strictly outside $\C$, and $f \in \inte\triangle(w',b,c)$. In particular,
    $f$ is a front vertex of $\mV$ from $w'$.
\end{lemma}

\begin{proof}
    For $w'=w$, the definition of a front vertex gives
    $f\in\inte\triangle(w,b,c)$. Any other witness $w'\in\W(\{b,c\})$
    lies on none of the lines $fb,fc,wf$ by \condref[]{general-position};
    since $w$ misses $f$, \lemref{cone} puts $w'$ in $\oppcone{f}{b}{c}$,
    so $f\in\inte\triangle(w',b,c)$. If $w'\in\C$, convexity of $\C$
    would give $\inte\triangle(w',b,c)\subseteq\inte\C$, contradicting
    $f\in\partial\C$.
\end{proof}

Conversely, placing a hull vertex $v$ inside $\triangle(p, s_1, s_2)$ for
\emph{any} two points $s_1, s_2 \in \C \setminus \{v\}$---not necessarily
the hull neighbors of $v$---already forces $v$ to be a front vertex from
$p$:

\begin{lemma}
    \lemlab{front-restoration}
    Let $v \in \mV$ with $\cardin{\mV} \ge 3$, let $s_1, s_2 \in \CHX{\mV}
    \setminus \{v\}$ be distinct, and let $p \in \R^2$ lie on no line
    through two points of $\mV$. If $v \in \inte\triangle(p, s_1, s_2)$,
    then $v$ is a front vertex of $\mV$ from $p$.
\end{lemma}

\begin{proof}
    Since $v \in \inte\triangle(p, s_1, s_2)$, we can write $p - v =
    -\alpha(s_1 - v) - \beta(s_2 - v)$ with $\alpha, \beta > 0$. For
    either edge $e$ of $\CHX{\mV}$ incident to $v$, let $L$ be a linear
    functional vanishing on the direction of $e$ and strictly positive on
    $\Set{q - v}{q \in \CHX{\mV} \setminus e}$. Because $s_1, s_2 \in
    \CHX{\mV} \setminus \{v\}$ and $p$ is not collinear with $e$, we have $L(s_1
    - v) \ge 0$ and $L(s_2 - v) \ge 0$, not both zero, so $L(p - v) < 0$.
    Thus both edges of $\CHX{\mV}$ incident to $v$ face $p$.
\end{proof}

\subsection{The certified block}
\seclab{certified-block}

\begin{definition}
    \deflab{certified}
    A hull vertex $f \in \mV$ is \emphi{certified} if some witness $w \in
    \W$ sees both neighbors of $f$ in $\mV$, misses $f$, and has $f$ as a
    front vertex of $\mV$ from $w$ (\figref{certified-vertex}). Let $\mH
    \subseteq \mV$ denote the set of certified hull vertices, and put
    \[
        \mD = \mV \setminus \mH
        \qquad\text{and}\qquad
        \mM = \Sin \cup \mH = \mS \setminus \mD
    \]
    for the set of uncertified hull vertices and the set of remaining
    targets, respectively.
\end{definition}

We now show that $\mH$ forms a single contiguous block along
$\partial\C$. Write
\[
    w_\star = w_\mS
    \qquad\text{and}\qquad
    \missw{b} = w_{\mS\setminus\{b\}} \quad (b \in \mS)
\]
for the universal witness seeing every target and the single-miss witness
missing only $b$. For every $f\in \mH$, both $w_\star$ and $\missw{f}$
see the two hull neighbors of $f$ in $\mV$, so \lemref{hidden-front}
implies that $f$ is a front vertex of $\mV$ from both $w_\star$ and
$\missw{f}$. Consequently, the line $\missw{f} f$ separates the two
neighbors of $f$ in $\mV$ and contains no other target of $\mS$ by
\condref[]{general-position}.

\begin{lemma}
    \lemlab{side-corners}
    Let $f \in \mH$, and let $u, v \in \mS \setminus \{f\}$ lie on opposite
    open sides of the line $\missw{f} f$. Then $f \in \inte\triangle(w, u, v)$
    for every witness $w \in \W(\{u, v\})$.
\end{lemma}

\begin{proof}
    Let $f_-, f_+$ be the neighbors of $f$ in $\mV$. Since $f$ is a front
    vertex of $\mV$ from $\missw{f}$, the lines $f_- f$ and $f f_+$ separate
    $\missw{f}$ from $\CHX{\mV} \setminus (f_- f)$ and $\CHX{\mV} \setminus (f
    f_+)$, respectively. Because $f$ is an extreme point of $\CHX{\mV}$ and
    $\missw{f} f$ meets $f_- f$ and $f f_+$ only at $f$, the open segment $uv
    \subset \CHX{\mV}$ crosses $\missw{f} f$ at a point $y \in \CHX{\mV} \setminus
    (f_- f \cup f f_+)$ strictly on the side of $f$ opposite $\missw{f}$. Hence
    $f \in \inte\triangle(\missw{f}, u, v)$. Since $\missw{f}$
    sees $u, v$ and misses $f$, \lemref{cone} and \condref[]{general-position}
    give $f \in \inte\triangle(w, u, v)$ for all $w \in \W(\{u, v\})$
    (immediately when $w = \missw{f}$).
\end{proof}

When $\mH \neq \emptyset$, every $f \in \mH$ is a front vertex of $\mV$
from $w_\star$, so $w_\star \notin \CHX{\mV}$ and all of $\mH$ lies in the
interior of the near chain $N_{w_\star}$ of $\mV$ from $w_\star$
(\figref{certified-block}).

\begin{figure}[htb]
    \centering
    \includegraphics[width=0.76\textwidth]{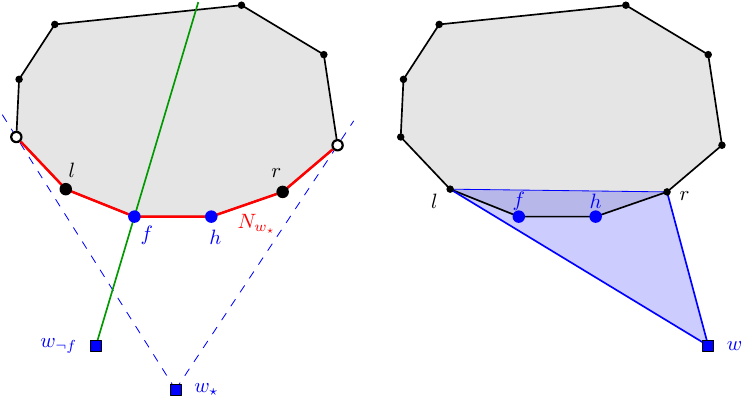}
    \caption{Left: the certified block $\mH=\{f,h\}$ (blue) inside the
        near chain $N_{w_\star}$ (red), flanked by $l,r\in\mD$ (black) and
        tangent endpoints (hollow circles). The green line $\missw{f}f$
        separates the chain vertices before $f$ from those after $f$.
        Right: every witness $w$ seeing $l$ and $r$ contains $\mH$ inside
        $\inte\triangle(w,l,r)$.}
    \figlab{certified-block}
\end{figure}

\begin{lemma}
    \lemlab{separation}
    For every $f\in \mH$, the line $\missw{f} f$ separates the vertices preceding
    $f$ along $N_{w_\star}$ from the vertices following $f$ along $N_{w_\star}$.
\end{lemma}

\begin{proof}
    The two hull neighbors of $f$ immediately precede and follow $f$ along
    $N_{w_\star}$ and lie on opposite sides of $\missw{f} f$. If two
    consecutive vertices $v, v'$ along $N_{w_\star}$, both preceding $f$
    or both following $f$, lay on opposite sides of $\missw{f} f$, then
    \lemref{side-corners} applied to $w = w_\star$ would give $f \in
    \inte\triangle(w_\star, v, v')$, placing the ray $w_\star f$ strictly
    between $w_\star v$ and $w_\star v'$ and contradicting strict angular
    monotonicity of $N_{w_\star}$ around $w_\star$.
\end{proof}

\begin{proposition}
    \proplab{certified-structure}
    Suppose $\cardin{\mS} \ge 3$ and $\mH \neq \emptyset$. Then:
    \begin{compactenumi}
    \item The vertices of $\mH$ form a single contiguous block along
        $\partial\CHX{\mV}$ inside $N_{w_\star}$. The two hull vertices $l, r \in \mD
        \cap N_{w_\star}$ immediately adjacent to $\mH$ are distinct, so
        $\cardin{\mD} \ge 2$.
    \item $\mH \subset \inte\triangle(w, l, r)$ for every witness $w \in
        \W(\{l, r\})$.
    \item $\cardin{\mM} = \cardin{\Sin} + \cardin{\mH} \le 5$.
    \end{compactenumi}
\end{proposition}

\begin{proof}
    Traverse $N_{w_\star}$ from one tangent endpoint to the other. Every
    certified vertex is an internal vertex of $N_{w_\star}$, whereas both
    tangent endpoints are uncertified and distinct. Let $h_-$ and $h_+$
    be the first and last vertices of $\mH$ along $N_{w_\star}$ (with
    $h_- = h_+$ if $\cardin{\mH} = 1$), let $l$ immediately precede $h_-$,
    and let $r$ immediately follow $h_+$. Then $l$ and $r$ exist, are
    distinct, and belong to $\mD \cap N_{w_\star}$. Once we exclude
    uncertified vertices between $h_-$ and $h_+$ along $N_{w_\star}$,
    $l$ and $r$ will be the two hull neighbors of the block $\mH$.

    Fix any $h\in \mH$. Because $l$ precedes $h$ and $r$ follows $h$ along
    $N_{w_\star}$, \lemref{separation} places $l$ and $r$ on opposite open
    sides of $\missw{h}h$. By \lemref{side-corners}, every witness $w \in
    \W(\{l,r\})$ satisfies $h\in\inte\triangle(w,l,r)$, proving~(ii). In
    particular, $\missw{h}$ sees $l,r$ and misses
    $h\in\inte\triangle(\missw{h},l,r)$. Thus taking $b_h=l$, $c_h=r$, and
    $w^\circ_h=\missw{h}$ satisfies the hypotheses of
    \lemref{hidden-targets} for $\mT=\mH$, proving~(iii).

    To complete~(i), suppose $\cardin{\mH}\ge 2$ and let $t$ lie strictly
    between $h_-$ and $h_+$ along $N_{w_\star}$. The single-miss witness
    $w=\missw{t}$ sees both $l$ and $r$, so part~(ii) gives
    $h_-, h_+ \in \inte\triangle(w,l,r)$. By the front-vertex restoration
    lemma (\lemref{front-restoration}), $h_-$ and $h_+$ are front vertices of $\mV$ from $w$, and hence
    internal vertices of the near chain $N_w$. Deleting $h_-$ and $h_+$
    splits $\partial\CHX{\mV}$ into two open hull arcs: the arc inside
    $N_{w_\star}$ containing $t$, and the complementary arc containing
    both $l$ and $r$. Because $N_w$ is a contiguous chain along
    $\partial\CHX{\mV}$, its subchain between $h_-$ and $h_+$ must follow
    one of these two arcs. If it followed the arc through $l$ and $r$,
    angular monotonicity of $N_w$ around $w$ would place the rays $wl$
    and $wr$ between $wh_-$ and $wh_+$, contradicting
    $h_-\in\inte\triangle(w,l,r)$ (which places $wh_-$ strictly between
    $wl$ and $wr$). Thus $N_w$ follows the arc containing $t$, making $t$
    an internal vertex of $N_w$ and hence a front vertex of $\mV$ from
    $\missw{t}$. Since $\missw{t}$ misses $t$ and sees both hull neighbors
    of $t$, we conclude $t\in\mH$.
\end{proof}

When $\mH = \emptyset$, \lemref{hidden-targets} with $\mT=\emptyset$
directly gives $\cardin{\mM} = \cardin{\Sin} \le 5$ and $\cardin{\mD} =
\cardin{\mV} \ge 3$.

\subsection{Exit edges}
\seclab{exit-edges}

Shattering supplies a witness $w_\mD$ seeing precisely $\mD$ and missing
every target in $\mM$. When $\cardin{\mD}\ge 3$, shooting rays from
$w_\mD$ through the missed targets assigns each $y \in \mM$ to an edge of
$\CHX{\mD}$ (\figref{vcdim6-overview}). Note that when $\mH\ne\emptyset$,
deleting the contiguous block $\mH$ from $\partial\CHX{\mV}$ makes its
flanking vertices $l$ and $r$ adjacent along $\partial\CHX{\mD}$, so $lr$
is an edge of $\CHX{\mD}$.

\begin{lemma}
    \lemlab{exit-edge}
    Suppose $\nu=\cardin{\mD}\ge 3$, and label the vertices of $\mD$
    counterclockwise around $\partial\CHX{\mD}$ from $0$ to $\nu-1$. For
    each $y\in\mM$, there is a unique edge $u_yv_y$ of $\CHX{\mD}$ not
    facing $w_\mD$ such that $y\in\inte\triangle(w_\mD,u_y,v_y)$; we call
    $e(y)=u_yv_y$ the \emphi{exit edge} of $y$. Every witness $w \in
    \W(\{u_y,v_y\})$ satisfies $y\in\inte\triangle(w,u_y,v_y)$. Moreover,
    if $\mH\ne\emptyset$, then $lr$ faces $w_\mD$ and therefore cannot be
    an exit edge.
\end{lemma}

\begin{proof}
    When $\mH=\emptyset$, we have $\mD=\mV$ and $y\in\Sin\subset\inte\CHX{\mD}$.
    When $\mH\ne\emptyset$, \propref{certified-structure} (ii) applied to
    $w_\mD \in \W(\{l,r\})$ puts $\mH\subset\inte\triangle(w_\mD,l,r)$, so
    $\CHX{\mV}\subseteq\CHX{\mD\cup\{w_\mD\}}$ in both cases; moreover,
    since the chord $lr$ of $\CHX{\mV}$ separates $\mH$ (and hence
    $w_\mD$) from $\mD\setminus\{l,r\}$, the edge $lr$ of $\CHX{\mD}$
    faces $w_\mD$.

    Every point of $\CHX{\mD\cup\{w_\mD\}}$ lies on a segment from $w_\mD$
    to $\CHX{\mD}$, so the ray from $w_\mD$ through $y\in\mM$ meets
    $\CHX{\mD}$ at or beyond $y$. By \condref[]{general-position}, this
    ray misses every vertex of $\mD$. Its last point in $\CHX{\mD}$
    therefore lies in the relative interior of a unique edge $u_yv_y$ of
    $\CHX{\mD}$, and $u_yv_y$ does not face $w_\mD$. Because $y$ lies off
    all lines spanned by $\mD$ by \condref[]{general-position}, we have
    $y\in\inte\triangle(w_\mD,u_y,v_y)$. Conversely, any edge of
    $\CHX{\mD}$ not facing $w_\mD$ whose triangle with $w_\mD$ contains
    $y$ must contain the ray's exit point from $\CHX{\mD}$, proving
    uniqueness.

    Finally, $w_\mD$ sees $u_y,v_y$ and misses $y$. Any other witness
    $w\in\W(\{u_y,v_y\})$ avoids the lines $yu_y,yv_y,yw_\mD$ by
    \condref[]{general-position}, so \lemref{cone} gives
    $y\in\inte\triangle(w,u_y,v_y)$.
\end{proof}

\subsection{The \texorpdfstring{$74$}{74} hull cases for seven targets}\seclab{case-enumeration}

Now specialize to $\cardin{\mS}=7$. By \propref{certified-structure} (iii)
when $\mH\ne\emptyset$ and \lemref{hidden-targets} when $\mH=\emptyset$,
the missed set $\mM=\Sin\cup\mH$ satisfies $\cardin{\mM}\le 5$, so
$\cardin{\mD} = 7 - \cardin{\mM} \ge 2$. We classify $(P,\mS,\W)$ up to
dihedral symmetries of the hull as follows.

\paragraph{All seven targets on the hull ($\Sin=\emptyset$).}
Here $\cardin{\mV}=7$, $\mM=\mH$, and $h=\cardin{\mH}\in\{0,1,\ldots,5\}$.
Cyclic relabeling around $\partial\CHX{\mV}$ places any nonempty
contiguous block $\mH$ in a canonical position, yielding six cases
indexed by $(\cardin{\mD},\cardin{\mH}) = (7-h, h)$ without exit-edge
labels.

\paragraph{At least one interior target ($\Sin\ne\emptyset$).}
Here $\cardin{\Sin} = 7 - \cardin{\mD} - \cardin{\mH} \ge 1$, so
$3 \le \cardin{\mD} + \cardin{\mH} = \cardin{\mV} \le 6$.
\begin{compactitem}
\item When $\nu = \cardin{\mD} \ge 3$, label $\mD = \{0,\dots,\nu-1\}$
    counterclockwise around $\partial\CHX{\mD}$, placing the pocket edge
    $lr$ at $(\nu-1)0$ whenever $h = \cardin{\mH} \ge 1$. By
    \lemref{exit-edge}, each target in $\mM$ takes its exit edge from the
    $\nu$ edges of $\CHX{\mD}$ if $h=0$, and from the $\nu-1$ edges other
    than $lr$ if $h\ge 1$. A labeling records the ordered $h$-tuple of
    exit edges of $\mH$ (in counterclockwise order from $l$ to $r$)
    together with the unordered multiset of $7-\nu-h$ exit edges of
    $\Sin$. We identify two labelings when they lie in the same orbit
    under the dihedral group of $\CHX{\mD}$ ($2\nu$ rotations and
    reflections when $h=0$, and the reflection fixing $lr$ while reversing
    the ordered $h$-tuple of $\mH$ when $h\ge 1$).
\item When $\cardin{\mD}=2$, $\CHX{\mD}$ is a single segment without fan
    triangles; since $\cardin{\mV}=\cardin{\mD}+\cardin{\mH}\ge 3$ and
    $\cardin{\Sin}\ge 1$, the size $h=\cardin{\mH}\in\{1,2,3,4\}$ alone
    specifies the case, giving four cases.
\end{compactitem}

\begin{table}[htb]
    \centering
    \begin{tabular}{@{}cccc@{}}
        \toprule
        $\cardin{\mD}$ & $\cardin{\mM}$ & Possible $\cardin{\mH}$ &
        Cases\\
        \midrule
        6&1&0&1\\
        5&2&0, 1&11\\
        4&3&0, 1, 2&29\\
        3&4&0, 1, 2, 3&23\\
        2&5&1, 2, 3, 4&4\\
        \midrule
        &&Total&68\\
        \bottomrule
    \end{tabular}
    \caption{Counts of hull cases with $\Sin\ne\emptyset$ modulo dihedral
        symmetries of $\CHX{\mD}$. For example, at $\cardin{\mD}=6$ the
        single interior target's exit edge rotates to $01$ ($1$ case); at
        $\cardin{\mD}=5$, two interior exit edges on a pentagon form $3$
        dihedral orbits for $\cardin{\mH}=0$, while for $\cardin{\mH}=1$
        the $4^2=16$ ordered pairs of non-$lr$ exit edges pair up under
        reflection into $8$ orbits ($3+8=11$ cases).}
    \tablab{coverage}
\end{table}

\tabref{coverage} lists the resulting $68$ cases with $\Sin\ne\emptyset$.
Together with the six all-hull cases, every shattered seven-target
configuration must fall into one of $68+6=74$ hull cases. We next derive
propositional constraints that refute all $74$ cases.

\section{Visibility constraints and clause families}
\seclab{constraints}
Fix a hull case and consider a candidate placement of the targets and
witnesses. The prescribed links from each $w_\mT$ to $\mT$ enclose
regions inside $P$ (\figref{forced-sightline}); whenever those enclosed
regions force $w_\mT$ to see a target outside $\mT$, the placement is
impossible. \secref{enclosure-links} derives forced visibilities from
convex quadrilaterals, reflex darts, and iterated path/triangle enclosure
(\lemref{four-cycle-enclosure} and \lemref{iterated-enclosure}).
\secref{half-plane} proves a half-plane and angular-ordering constraint
induced by a witness missing two targets (\lemref{half-plane}).
\secref{chirotope} sets up the Boolean encoding of orientations and
links, and \secref{soundness} states all $11$ clause families and proves
their soundness (\thmref{chirotope-sound}).

\subsection{How enclosure forces visibility}
\seclab{enclosure-links}

When two witnesses $w_1, w_2$ both see two targets $g, b$, the four links
form a closed polygonal $4$-cycle $C_4 = w_1 - g - w_2 - b - w_1$ in $P$.
Whenever $C_4$ does not cross itself, it bounds either a convex
quadrilateral or a reflex dart inside $P$, forcing visibility across its
convex subregions (\figref{enclosure-dart}).

\begin{figure}[t]
    \centering
    \includegraphics[width=0.92\textwidth]{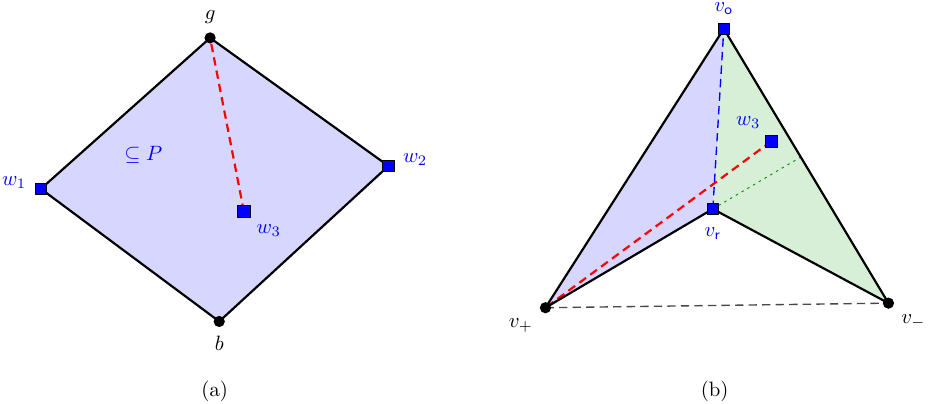}
    \caption{Regions enclosed in $P$ by a non-crossing $4$-cycle of links.
        (a)~A strictly convex quadrilateral $\CHX{\{w_1,g,w_2,b\}}\subseteq
        P$, forcing an interior witness $w_3$ to see $g$ (red dashed
        link). (b)~A reflex dart $\cD = \triangle(v_{\mr},v_+,v_{\mo})\cup
        \triangle(v_{\mr},v_{\mo},v_-)\subseteq P$ with inward corner
        $v_{\mr}$. Extending $v_+v_{\mr}$ (dotted green) cuts off a
        convex subpolygon on the $v_{\mo}$-side containing both $v_+$ and
        $w_3$, forcing the red dashed link $v_+w_3$.}
    \figlab{enclosure-dart}
\end{figure}

\begin{lemma}
    \lemlab{four-cycle-enclosure}
    Let $w_1, g, w_2, b \in P$ be four distinct points with no three
    collinear such that $w_1$ and $w_2$ both see $g$ and $b$, and let
    $C_4 = (w_1, g, w_2, b)$ be the closed polygonal $4$-cycle $w_1 - g
    - w_2 - b - w_1$.
    \begin{compactenumi}
    \item If $C_4$ bounds a strictly convex quadrilateral---that is, its
        four consecutive turns $(w_1,g,w_2)$, $(g,w_2,b)$, $(w_2,b,w_1)$,
        and $(b,w_1,g)$ have the same orientation sign (\figref{enclosure-dart} (a))---then
        $\CHX{\{w_1,g,w_2,b\}} \subseteq P$, and every two points in
        $\CHX{\{w_1,g,w_2,b\}}$ see each other.
    \item Suppose $C_4$ forms a reflex dart, meaning that one vertex
        $v_{\mr} \in \{w_1, g, w_2, b\}$ lies in $\inte\triangle(v_+,
        v_{\mo}, v_-)$, where $(v_{\mr}, v_+, v_{\mo}, v_-)$ is a cyclic
        shift of $C_4$ so $v_+, v_-$ are adjacent to $v_{\mr}$ and
        $v_{\mo}$ is opposite $v_{\mr}$ (\figref{enclosure-dart} (b)).
        Then the dart region
        \[
            \cD = \triangle(v_{\mr},v_+,v_{\mo})\cup
            \triangle(v_{\mr},v_{\mo},v_-)
            = \triangle(v_+,v_{\mo},v_-) \setminus
            \pth{\inte\triangle(v_{\mr},v_+,v_-) \cup
            (v_+v_-\setminus\{v_+,v_-\})}
        \]
        is contained in $P$, and:
        \begin{compactenuma}
        \item both $v_{\mr}$ and $v_{\mo}$ see every point of $\cD$,
        \item every point in $\inte\triangle(v_+, v_{\mo}, v_-)$ on the
            $v_{\mo}$-side of the line $v_{\mr} v_+$ lies in $\cD$ and
            sees $v_+$ as well as every other point in
            $\inte\triangle(v_+, v_{\mo}, v_-)$ on the $v_{\mo}$-side of $v_{\mr}
            v_+$, and
        \item symmetrically, every point in $\inte\triangle(v_+, v_{\mo},
            v_-)$ on the $v_{\mo}$-side of the line $v_{\mr} v_-$ lies in
            $\cD$ and sees $v_-$ as well as every other point in
            $\inte\triangle(v_+, v_{\mo}, v_-)$ on the $v_{\mo}$-side of $v_{\mr} v_-$.
        \end{compactenuma}
    \end{compactenumi}
\end{lemma}

\begin{proof}
    All four edges of $C_4$ lie in $P$. In~(i), $C_4$ is a simple closed
    polygonal curve enclosing $\CHX{\{w_1, g, w_2, b\}}$, so
    \lemref{enclosure} (i) gives $\CHX{\{w_1, g, w_2, b\}} \subseteq P$ and
    convexity gives every stated visibility.

    In~(ii), since $v_{\mr} \in \inte\triangle(v_+, v_{\mo}, v_-)$, the simple
    closed curve $v_{\mr} - v_+ - v_{\mo} - v_- - v_{\mr} \subset P$ encloses
    $\cD = \triangle(v_{\mr}, v_+, v_{\mo}) \cup \triangle(v_{\mr}, v_{\mo},
    v_-)$, so $\cD \subseteq P$ by \lemref{enclosure} (i). Both
    constituent triangles are convex and contain $v_{\mr}$ and $v_{\mo}$,
    giving~(a). For~(b), because $v_{\mr} \in \inte\triangle(v_+, v_{\mo}, v_-)$,
    the line $v_{\mr} v_+$ separates $v_{\mo}$ from $v_-$. Hence $\triangle(v_{\mr},
    v_+, v_-) \setminus (v_{\mr} v_+)$ lies entirely on the $v_-$-side of
    $v_{\mr} v_+$. Intersecting $\triangle(v_+, v_{\mo}, v_-)$ with the closed
    $v_{\mo}$-half-plane $H_{v_{\mo}}$ of $v_{\mr} v_+$ therefore avoids
    $\inte\triangle(v_{\mr}, v_+, v_-) \cup (v_+ v_- \setminus \{v_+\})$ and
    yields a convex polygon $K_+ = \triangle(v_+, v_{\mo}, v_-) \cap H_{v_{\mo}}
    \subseteq \cD \subseteq P$ containing $v_+$. Convexity of
    $K_+$ gives~(b), and~(c) is symmetric.
\end{proof}

Applying two-path enclosure and triangle containment repeatedly from a
known graph of links yields only true links:

\begin{lemma}
    \lemlab{iterated-enclosure}
    Let $\mX \subset P$ be a finite point set, and let $G_0$ be a graph on
    $\mX$ whose edges all represent visible pairs in $P$. Define $G_{\tau+1}$
    from $G_\tau$ ($\tau \ge 0$) by adding every pair $uv \in \binom{\mX}{2}$
    satisfying at least one of:
    \begin{conditions}{E}
    \item\condlab{opposite-paths} there exist two paths $\pi_1, \pi_2$ from $u$ to
        $v$ in $G_\tau$, each with at least two edges, with all internal vertices of
        $\pi_1$ on one open side of $uv$ and all internal vertices of $\pi_2$ on the other, or
    \item\condlab{triangle-links} there exists a triangle $\{g, b, c\}$ in $G_\tau$
        such that $u \in \inte\triangle(g, b, c)$ and $v \in \{g, b, c\}
        \cup \inte\triangle(g, b, c)$.
    \end{conditions}
    Then every edge of $G_\tau$ ($\tau \ge 0$) is a visible pair in $P$. In
    particular, if $g, b \in \mS$ are distinct and two witnesses $w_1,w_2$
    that see both targets lie on opposite sides of the line $gb$, then
    $gb \subseteq P$, and hence $\triangle(w_3, g, b) \subseteq P$ for
    every witness $w_3$ that sees both $g$ and $b$.
\end{lemma}

\begin{proof}
    Induct on $\tau \ge 0$. The base $\tau = 0$ holds by assumption. Given that
    all edges of $G_\tau$ lie in $P$, every pair $uv$ added by \condref[]{opposite-paths}
    satisfies $uv \subseteq P$ by \lemref{enclosure} (ii), and every pair
    added by \condref[]{triangle-links} lies in $\triangle(g, b, c) \subseteq P$ by
    \lemref{enclosure} (i). Applying \condref[]{opposite-paths} to $g - w_1 - b$ and $g
    - w_2 - b$ gives $gb \subseteq P$, after which \lemref{enclosure} (i)
    gives $\triangle(w_3, g, b) \subseteq P$ (so \condref[]{triangle-links} applies to
    $\triangle(w_3, g, b)$).
\end{proof}

\subsection{The half-plane lemma}
\seclab{half-plane}

Gilbers and Klein~\cite{GK14} used half-plane tests to bound targets along
a convex chain. We extract a local orientation version involving two
witnesses $w_0, w_1$ and three targets $g, p, c$ (\figref{half-plane-frame}).
Normalize coordinates so that $w_0=(0,0)$ sees $p=(0,1)$ and misses $g$
and $c$, while $w_1=(1,0)$ sees all three targets. Suppose $g$ lies in
the open first quadrant, $c$ lies in the open second quadrant, and the
segment $gc$ passes below $p$ so that $gc$ crosses $w_0p$. Suppose
further that either $w_0$ sees $w_1$ directly, or a common visible target
$e$ to the right of $w_0g$ forces the shaded subpolygon $\Delta_g$ of
$\triangle(w_0,w_1,p)$ above $w_0g$ to lie inside $P$. Then the link
$w_1c$ crosses $\Delta_g$ from $d_0\in w_0g$ to $z\in w_0p$, forming a
barrier in $P$: every witness $w\in\W(\{g,c\})$ must lie below $w_0g$ and
to the right of $w_0p$, and viewed from $p$, witnesses in $\W(\{g,c\})$
that see $p$ precede both $g$ and every witness in $\W(\{g,c\})$ that
misses $p$. Clause family \condref[]{half-plane} records these
implications.

\begin{figure}[htb]
    \centering
    \includegraphics[width=.43\textwidth]{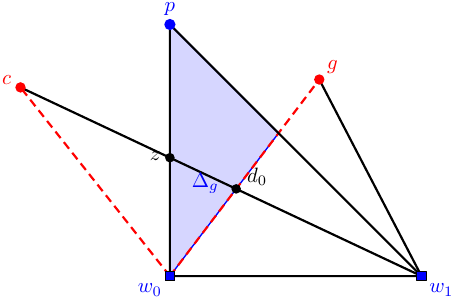}
    \caption{Normalized frame with $(w_0,w_1,p)$ oriented
        counterclockwise ($\orient(w_0,w_1,p)>0$):
        $w_0=(0,0)$, $w_1=(1,0)$, and $p=(0,1)$. Red dashed segments mark
        the blocked sightlines $w_0c$ and $w_0g$. The shaded region
        $\Delta_g\subseteq P$ is the part of $\triangle(w_0,w_1,p)$ on the
        $p$-side of $w_0g$ (containing $w_0d_0$), and the link $w_1c$
        crosses $\partial\Delta_g$ at $d_0\in w_0g$ and $z\in w_0p$. The
        witnesses $w_0$ and $w_1$ need not see each other directly.}
    \figlab{half-plane-frame}
\end{figure}

\begin{lemma}
    \lemlab{half-plane}
    Let $w_0, w_1 \in \W$ be distinct, let $g, p, c \in \mS$ be distinct,
    and set $\sigma=\sgn\orient(w_0,\allowbreak w_1,\allowbreak p)\allowbreak\in\{-1,+1\}$.
    Assume the following conditions:
    \begin{compactenumi}
    \item $w_0$ sees $p$ and misses $g$ and $c$, while $w_1$ sees $g, p, c$,
    \item $\sgn\orient(w_0, w_1, g) = \sgn\orient(w_0, w_1, c) = \sgn\orient(w_0,
        g, p) = \sgn\orient(w_0, p, c) = \sigma$,
    \item $\sgn\orient(g, p, c) = \sigma$ (equivalently, $\sgn\orient(g,
        c, w_0) = -\sgn\orient(g,\allowbreak c,\allowbreak p) = \sigma$, so $w_0p$ and $gc$ cross), and
    \item either $w_0$ sees $w_1$, or there exists $e \in \vis(w_0) \cap
        \vis(w_1) \cap \mS$ satisfying $\sgn\orient(w_0, w_1, e) = \sgn\orient(w_0,
        e, g) = \sigma$.
    \end{compactenumi}
    Then every $w \in \W(\{g, c\})$ satisfies $\sgn\orient(w_0, g, w) =
    -\sigma$ and $\sgn\orient(w_0, p, w) = -\sigma$, and every $w \in
    \W(\{g,\allowbreak c,\allowbreak p\})$ satisfies $\sgn\orient(p,\allowbreak w,\allowbreak g) = \sigma$ and
    $\sgn\orient(p, w, w') = \sigma$ for all $w' \in \witnessfamily{\{g, c\}}{\{p\}}$.
\end{lemma}

\begin{proof}
    Apply the affine transformation mapping $w_0 \mapsto (0, 0)$, $w_1
    \mapsto (1, 0)$, and $p \mapsto (0, 1)$, so that normalized
    orientations
    \[
        \normorient(x, y, z) = \frac{\sigma}{\abs{\orient(w_0, w_1,
        p)}}\,\orient(x, y, z)
    \]
    preserve orientation signs up to the global factor $\sigma$. Write
    $g = (g_1, g_2)$ and $c = (c_1, c_2)$ (\figref{half-plane-frame});
    condition~(ii) translates directly to
    \begin{equation}
        \eqlab{signs}
        g_1 > 0, \qquad g_2 > 0, \qquad c_1 < 0, \qquad c_2 > 0.
    \end{equation}
    Put $\Delta_0 = \triangle(w_0, w_1, p)$ and $\Delta_g = \Set{y \in
    \Delta_0}{\normorient(w_0, g, y) \ge 0}$. We prove each conclusion by
    contradiction: a witness on the forbidden side closes two link paths
    around a sightline missed by $w_0$ or $w'$.

    \begin{claim}
        \clmlab{Ta}
        $\Delta_g \subseteq P$.
    \end{claim}
    \begin{proof}
        If $w_0$ sees $w_1$, then $\Delta_0 \subseteq P$ by \lemref{enclosure} (i)
        since $w_0p, pw_1, w_0w_1 \subseteq P$. Otherwise let $e = (e_1, e_2) \in
        \mS$ be as in~(iv), so $e_2 > 0$, $e_1 g_2 - e_2 g_1 > 0$ (hence
        $e_1 > 0$ by \Eqref{signs}), and $e \notin pw_1$ since no three
        points of $\mS\cup\W$ are collinear by
        \condref[]{general-position}. The ray $w_0e$ meets $pw_1$ at
        $x^*$, and $\Delta_g \subseteq \triangle(w_0, x^*, p)$. If
        $e \notin \Delta_0$, then $x^* \in w_0e \subseteq P$, giving
        $\triangle(w_0, x^*, p) \subseteq P$ by \lemref{enclosure} (i). If
        $e \in \Delta_0$, the simple closed curve $w_0 - e - w_1 - p - w_0 \subset
        P$ encloses the reflex dart $\triangle(w_0, e, p) \cup \triangle(e,
        w_1, p) \supseteq \triangle(w_0, x^*, p)$ in $P$ (\lemref{enclosure} (i)).
    \end{proof}

    \begin{claim}
        \clmlab{coords}
        We have $g_1 + g_2 > 1$, $c_1 + c_2 < 1$, and $-1 < s_g < s_c$,
        where $s_g = (g_2 - 1)/g_1$ and $s_c = (c_2 - 1)/c_1$.
        Moreover, $w_1c$ crosses $w_0p$ at $z = (0, z_2)$ ($0 < z_2 < 1$) and
        crosses $w_0g$ at $d_0 \in \Delta_g$.
    \end{claim}
    \begin{proof}
        If $g$ lay in $\Delta_0$, the segment $w_0g$ would lie in
        $\Delta_g\subseteq P$ (\clmref{Ta}), contradicting $g\notin\vis(w_0)$.
        Since $g_1,g_2>0$ by \Eqref{signs}, $g$ lies strictly beyond $pw_1$,
        giving $g_1+g_2>1$ and hence $s_g=(g_2-1)/g_1 > -1$.
        Condition~(iii) gives
        \[
            0<\normorient(g,p,c)=g_1c_1(s_g-s_c).
        \]
        Since $g_1c_1<0$ by \Eqref{signs}, we obtain $s_c>s_g>-1$, and
        multiplying $s_c>-1$ by $c_1<0$ gives $c_1+c_2<1$.

        The segment $w_1c$ crosses the vertical axis at height
        $z_2=c_2/(1-c_1)$, which satisfies $0 < z_2 < 1$ because $c_2 > 0$,
        $c_1 < 0$, and $c_1 + c_2 < 1$; thus $z\in w_0p$. Moreover,
        $w_1$ and $z$ lie on opposite sides of $w_0g$, since
        $\normorient(w_0,g,w_1)=-g_2<0$ and
        $\normorient(w_0,g,z)=g_1z_2>0$. The open segment $w_1z \subset
        \Delta_0$ lies in the open first quadrant and therefore meets the
        open ray from $w_0$ through $g$ at a point $d_0\in\Delta_g$; since
        $g\notin\Delta_0$, $d_0$ lies strictly between $w_0$ and $g$.
    \end{proof}

    \begin{claim}
        \clmlab{witness-pos}
        For a point $r=(\xi_r,\eta_r)$ with $\xi_r\ne0$, put
        $s(r)=(\eta_r-1)/\xi_r$, the slope from $p$ to $r$. Every witness
        $w=(\xi,\eta)\in\W(\{g,c\})$ satisfies:
        \begin{compactenuma}
        \item $\normorient(w_0, g, w) < 0$,
        \item $\xi > 0$, i.e., $\normorient(w_0, p, w) < 0$, and
        \item if $w \in \W(\{g, c, p\})$, then $s(w)<s_g$, equivalently
            $\normorient(p,w,g)>0$.
        \end{compactenuma}
    \end{claim}
    \begin{proof}
        (a)~The points $w_1$ and $c$ lie on opposite sides of $w_0g$,
        since $\normorient(w_0,\allowbreak g,\allowbreak w_1)=-g_2<0$ and
        $\normorient(w_0,g,c)=g_1c_2-g_2c_1>0$ by \Eqref{signs}. If $w$
        lay on the $c$-side of $w_0g$, the link paths $g-w_1-d_0$ and
        $g-w-c-d_0$ (using $w_1d_0, cd_0 \subset w_1c$) would enclose
        $gd_0$; together with $w_0d_0\subseteq \Delta_g\subseteq P$
        (\clmref{Ta}), this would force $w_0$ to see $g$. Since
        \condref[]{general-position} excludes $w\in w_0g$, $w$ lies on
        the $w_1$-side.

        (b)~By \condref[]{general-position}, $\xi \neq 0$. Suppose $\xi <
        0$. Part~(a) gives $g_1 \eta < g_2 \xi < 0$, so $\eta < 0$ and the
        segment $gw$ crosses the vertical axis at $y = (0, y_2)$ with
        $y_2 = (g_1\eta - g_2\xi)/(g_1 - \xi) < 0$. The points $w,c$ lie
        to the left of the vertical line $yz$, while $g,w_1$ lie to its
        right, so the paths $y-w-c-z$ and $y-g-w_1-z$ enclose $yz\subseteq
        P$. Since $y_2<0<z_2$, we obtain $yw_0 \subset yz \subseteq P$.

        Next consider the line $cw_0$. Because $c_1<0<c_2$, $\xi,\eta<0$,
        and $y_2<0<z_2$, we have $\normorient(c,w_0,w) = -c_1\eta + c_2\xi
        < 0$, $\normorient(c,w_0,y) = -c_1y_2 < 0$, and
        $\normorient(c,w_0,z) = -c_1z_2 > 0$, placing $w,y$ on one open
        side of $cw_0$ and $z$ on the other. Using $wy\subset wg$,
        $yw_0\subset yz$, $cz\subset cw_1$, and $zw_0\subset pw_0$, the
        link paths $c-w-y-w_0$ and $c-z-w_0$ enclose $cw_0$,
        contradicting $c\notin\vis(w_0)$.

        (c)~We have $\normorient(p, w, g) = \xi g_1(s_g - s(w))$ with
        $\xi, g_1 > 0$. If $w = w_1$, then $s(w_1) = -1 < s_g$ by
        \clmref{coords}. If $w \neq w_1$ and $s(w) \ge s_g$ (strict by
        \condref[]{general-position}), then $p - w_1 - g$ and $p - w - g$
        enclose $pg$, so $\triangle(p, w_1, g) \subseteq P$ by
        \lemref{enclosure} (i); combined with $\Delta_g\subseteq P$
        (\clmref{Ta}), this puts the entire segment $w_0g$ inside $P$, a
        contradiction.
    \end{proof}

    \emph{Order the witnesses around $p$.} Let $w\in\W(\{g,c,p\})$ and
    $w'=(\xi',\eta')\in\witnessfamily{\{g,c\}}{\{p\}}$. By
    \clmref{witness-pos}, both $\xi$ and $\xi'$ are positive, and
    $s(w)<s_g$. Suppose for contradiction that $s(w')\le s(w)$; by
    \condref[]{general-position}, $p,w,w'$ are not collinear, so
    $s(w')<s(w)<s_g<s_c$. Then $\normorient(p,w',r)=\xi'\xi_r(s(r)-s(w'))$
    is positive for $r\in\{w,g\}$ and negative for $r=c$, while
    $\normorient(p,w',z)=\xi'(z_2-1)<0$. Thus $w,g$ lie on one open side
    of $pw'$ and $z,c$ lie on the other. Because $pz\subset pw_0$ and
    $zc\subset w_1c$, the link paths $p-w-g-w'$ and $p-z-c-w'$ lie in $P$
    and enclose $pw'$, contradicting $p\notin\vis(w')$. Hence
    $s(w')>s(w)$, i.e., $\normorient(p,w,w')>0$. Multiplying by $\sigma$
    restores all orientation signs in the original coordinates.
\end{proof}

\subsection{From links to orientation signs}
\seclab{chirotope}

\begin{definition}
    \deflab{chirotope}
    Let $\mX \subset \R^2$ be a finite set of points with no three
    collinear. The \emphi{chirotope} of $\mX$ is the alternating sign map
    $\chi$ on ordered triples of distinct elements of $\mX$ defined by
    \[
        \chi(p, q, r) = \sgn\orient(p, q, r) \in \{-1, +1\}.
    \]
\end{definition}

Fix a hull case and a witness subfamily $\W_0 \subseteq \W$, and put
$\mX=\mS\cup\W_0$. The propositional formula for the case uses three
types of Boolean variables:
\begin{compactitem}
\item \emph{Orientation variables:} one variable for each unordered
    triple of distinct points in $\mX$, recording $\chi(p,q,r)$ for one
    canonical ordering of the triple (with antisymmetry determining the
    other five orderings). Triples of hull vertices in $\mV$ are fixed to
    constants by their counterclockwise hull order.
\item \emph{Link variables:} one variable for each unordered pair of
    distinct points in $\mX$, true when the pair forms a link in $P$. For
    $w\in\W_0$ and $s\in\mS$, the link $ws$ is fixed to the truth constant
    prescribed by shattering; target--target and witness--witness pairs
    remain free variables.
\item \emph{Auxiliary variables:} gate, ordering, and coloring variables
    standing for intermediate geometric predicates (such as the
    antecedent of \lemref{half-plane} or the side/component coloring of a
    non-visible pair). The proof of \thmref{chirotope-sound} constructs
    their satisfying assignment in any true realization.
\end{compactitem}
Write $\mL(p,q,r)$ for the literal $\chi(p,q,r)=+1$, and write
$z\in\inte_\chi\triangle(p,q,r)$ as shorthand for
\[
    \chi(p,q,z)=\chi(q,r,z)=\chi(r,p,z)=\chi(p,q,r).
\]
For Euclidean orientation signs, $z\in\inte_\chi\triangle(p,q,r)$ holds
if and only if $z\in\inte\triangle(p,q,r)$.

\paragraph{An example.}
Suppose shattering requires $w_1,w_2,w\in\W_0$ to see $g,b\in\mS$ and
requires $w$ to miss $t\in\mS$ (\figref{forced-sightline}). If
$\mL(g,b,w_1)$ holds and $\mL(g,b,w_2)$ is false, $w_1$ and $w_2$ lie on
opposite sides of $gb$; the link paths $g-w_1-b$ and $g-w_2-b$ then
enclose $gb$, so \condref{paths} below forces the link variable of $gb$ to
be true. If $\mL(g,b,t)$, $\mL(b,w,t)$, and $\mL(w,g,t)$ also hold (which
by the four-point identity in \condref{plucker} also forces
$\mL(g,b,w)$), then $t\in\inte_\chi\triangle(g,b,w)$ inside a triangle of
three links, so \condref{triangles} forces the link $wt$, contradicting
$t\notin\vis(w)$. Together these clauses exclude the five-literal
conjunction
\[
    \mL(g,b,w_1) \;\wedge\; \neg\mL(g,b,w_2) \;\wedge\; \mL(g,b,t)
    \;\wedge\; \mL(b,w,t) \;\wedge\; \mL(w,g,t).
\]

\subsection{Soundness of the clause families}
\seclab{soundness}

Each of the $74$ case formulas is a conjunction of instances from the
$11$ clause families in \thmref{chirotope-sound}, instantiated on a
witness subfamily $\W_0$ containing the $29$ witnesses missing at most
two targets (the generator in the repository specifies $\W_0$ and the
exact subset of instances emitted per case; see \secref{certificates}).
\tabref{clause-families} summarizes the families. Recall the partition
$\mV = \mD \sqcup \mH$ into uncertified and certified hull vertices, the
interior targets $\Sin = \mS \setminus \mV$, and the targets
$\mM = \Sin \cup \mH$ missed by $w_\mD$.

\begin{table}[!htb]
    \centering
    \small
    \begin{tabular}{@{}lp{0.5\textwidth}l@{}}
        \toprule
        Family & Geometric content & Justified by\\
        \midrule
        \condref[]{plucker} & Grassmann--Pl\"ucker and four-point
        orientation identities & Knuth~\cite{Knu92}\\
        \condref[]{radial} & directions around a pivot are linearly
        ordered (used in two cases) & bilinearity\\
        \condref[]{hull-order} & hull vertices in counterclockwise order,
        interior targets inside & definition of $\mV$\\
        \condref[]{hull-obstruction} & no witness hides an uncertified hull
        vertex inside a triangle of seen targets & \lemref{cone},
        \lemref{front-restoration}\\
        \condref[]{exit-edge} & each missed target lies in its exit-edge
        triangle, with wedge separation & \lemref{exit-edge},
        \lemref{wedge-sep}\\
        \condref[]{certified-pocket} & the certified block lies in
        $\triangle(w,l,r)$ and is split by the lines $\missw{f}f$ &
        \propref{certified-structure}, \lemref{separation}\\
        \condref[]{half-plane} & half-plane constraints on witnesses
        seeing $g,c$ and their angular order around $p$ & \lemref{half-plane}\\
        \condref[]{paths} & a non-link segment is not enclosed by paths
        of links & \lemref{enclosure} (ii)\\
        \condref[]{triangles} & a triangle of links sees every point in
        its interior & \lemref{enclosure} (i)\\
        \condref[]{darts} & convex quadrilaterals and reflex darts of
        links (used in one case) & \lemref{four-cycle-enclosure}\\
        \condref[]{cuts} & saved cuts whose negation forces a forbidden
        link (used in one case) & \lemref{iterated-enclosure}\\
        \bottomrule
    \end{tabular}
    \caption{Summary of the $11$ clause families in \thmref{chirotope-sound}.}
    \tablab{clause-families}
\end{table}

\begin{theorem}
    \thmlab{chirotope-sound}
    Let $\mS \subset P$ be a shattered set of $\cardin{\mS} = 7$ points in a
    simple polygon $P$ under \condref[]{general-position}--\condref[]{witness-disks}, with $\mV =
    \partial\CHX{\mS} \cap \mS = \mD \sqcup \mH$ ($\nu = \cardin{\mD}$),
    $\Sin = \mS \setminus \mV$, and $\mM = \Sin \cup \mH$. Let $\W_0
    \subseteq \W$ and $\mX = \mS \cup \W_0$. Assign every orientation
    variable the Euclidean chirotope sign of its triple and every link
    variable the true visibility of its pair in $P$. Then the auxiliary
    variables can be assigned truth values so that every instance of the
    following families holds:
    \begin{conditions}{C}
    \item\condlab{plucker} \emph{Orientation identities:} For every
        ordered $5$-tuple $(p, g, b, c, d)$ of distinct points in $\mX$,
        the three products $\chi(p, g, b)\allowbreak\chi(p, c, d)$, $-\chi(p, g,
        c)\allowbreak\chi(p, b, d)$, and $\chi(p, g, d)\chi(p, b, c)$ cannot all be
        $+1$ and cannot all be $-1$. For distinct $p,q,r,z\in\mX$, if
        $\chi(p,q,z)=\chi(q,r,z)=\chi(r,p,z)$, then this common value
        equals $\chi(p,q,r)$.
    \item\condlab{radial} \emph{Radial order:} Fix $p\in\mX$ and a second
        point $r\in\mX$, and put $\epsilon_q=\chi(p,r,q)$ for each
        $q\in\mX\setminus\{p,r\}$. For distinct $q,s\in\mX\setminus\{p,r\}$,
        an auxiliary order variable records whether
        $\epsilon_q\epsilon_s\chi(p,q,s)=+1$. These order variables are
        transitive.
    \item\condlab{hull-order} \emph{Convex hull and interior containment:}
        Label the vertices $\mV=(v_0,\dots,v_{\cardin{\mV}-1})$ in counterclockwise
        order around $\partial\CHX{\mV}$, with $l$, the block $\mH$, and $r$
        consecutive when $\mH\ne\emptyset$. Then
        $\chi(v_\mu,v_\kappa,v_\lambda)=+1$ for all $0\le \mu<\kappa<\lambda<
        \cardin{\mV}$, and $\chi(v_\kappa, v_{\kappa+1}, g) = +1$ for every $g \in
        \Sin$ and every $0 \le \kappa < \cardin{\mV}$ (indices
        modulo $\cardin{\mV}$).
    \item\condlab{hull-obstruction} \emph{Uncertified hull obstruction:} For every
        $w \in \W_0$, every $v \in \mD \setminus \vis(w)$, and every
        distinct pair $s_1, s_2 \in \vis(w) \cap \mS$, we have $v \notin
        \inte_\chi\triangle(w, s_1, s_2)$.
    \item\condlab{exit-edge} \emph{Exit-edge cones and wedge separation (when
        $\nu \ge 3$):} For each $y \in \mM$
        with exit edge $u_y v_y$, every $w \in \W_0 \cap
        \W(\{u_y, v_y\})$ satisfies $y \in \inte_\chi\triangle(w, u_y,
        v_y)$, and for any distinct $w_1, w_2 \in
        \W_0 \cap \W(\{u_y, v_y, y\})$ and $w_0 \in \W_0 \cap \witnessfamily{\{u_y,
        v_y\}}{\{y\}}$, we do not have $\chi(y, w_1, w_2) = \chi(y, w_1,
        w_0) = \chi(y, w_0, w_2)$.
    \item\condlab{certified-pocket} \emph{Certified pocket geometry ($\mH \neq
        \emptyset$):} Every $f \in \mH$ with neighbors $f_-, f_+$ in $\mV$
        satisfies $f \in \inte_\chi\triangle(w, l, r)$ for all $w \in
        \W_0 \cap \W(\{l, r\})$ and
        $f \in \inte_\chi\triangle(w, f_-, f_+)$ for all $w \in \W_0 \cap
        \W(\{f_-, f_+\})$, with wedge
        separation around $f$ on both families as in \condref[]{exit-edge}.
        Moreover, whenever
        $\missw{f} \in \W_0$, the line $\missw{f} f$
        separates the $l$-side from the $r$-side of $(\mH \cup \{l, r\})
        \setminus \{f\}$, and any $u, v \in \mS
        \setminus \{f\}$ with $\chi(\missw{f}, f, u) = -\chi(\missw{f}, f, v)$
        satisfy $f \in \inte_\chi\triangle(w, u, v)$ for all $w \in \W_0
        \cap \W(\{u, v\})$.
    \item\condlab{half-plane} \emph{Half-plane constraints:} Let $w_1
        = w_\mS \in \W_0$, $w_0 \in \W_0 \setminus \{w_1\}$, and let
        distinct $g, p, c \in \mS$ satisfy $p \in \vis(w_0)$ and $g, c
        \notin \vis(w_0)$. Suppose that for some $\sigma \in \{-1, +1\}$,
        $\chi(w_0, w_1, g) = \chi(w_0, w_1, p) = \chi(w_0, w_1, c) = \chi(w_0, g, p) =
        \chi(w_0, p, c) = \chi(g, c, w_0) = -\chi(g, c, p) =
        \sigma$, and that either $w_0w_1$ is a link or some $e \in
        \vis(w_0) \cap \mS$ satisfies $\chi(w_0, w_1, e) = \chi(w_0, e, g) =
        \sigma$. Then every $w \in \W_0 \cap \W(\{g, c\})$ satisfies $\chi(w_0,
        g, w) = \chi(w_0, p, w) = -\sigma$, and every $w \in
        \W_0 \cap \W(\{g, c, p\})$ satisfies $\chi(p, w, g) = \sigma$ and
        $\chi(p, w, w') = \sigma$ for all $w' \in \W_0 \cap \witnessfamily{\{g, c\}}{\{p\}}$.
    \item\condlab{paths} \emph{Path separation:} For distinct $u, v \in \mX$
        that do not see each other, one of the two open sides of the line
        $uv$ is selected. In the graph of links induced on $u$, $v$, and
        the points of $\mX$ on the selected side, a two-coloring
        separates $u$ from $v$: linked points receive equal colors, and
        $u,v$ receive opposite colors. In particular, no point on the
        selected side is linked to both $u$ and $v$.
    \item\condlab{triangles} \emph{Triangle links:} If the three sides of a
        triangle with vertices in $\mX$ are links, every point of $\mX$
        in its interior is linked to each of its three vertices.
    \item\condlab{darts} \emph{Quadrilaterals and darts:} Let $w_1, w_2
        \in \W_0$ both see the targets $g, b$. If
        \lemref{four-cycle-enclosure} implies that a subregion determined
        by the $4$-cycle $(w_1, g, w_2, b)$ sees a target $s$, then every
        witness of $\W_0$ lying in that subregion sees $s$.
    \item\condlab{cuts} \emph{Saved cuts:} Let $C$ be a clause over the
        orientation variables such that, starting from the fixed hull
        signs of \condref[]{hull-order}, the orientation signs falsifying
        $C$, and the prescribed witness--target links, repeated
        application of rules \condref[]{opposite-paths} and
        \condref[]{triangle-links} of \lemref{iterated-enclosure} (using
        only those fixed signs) derives a link between some $w \in \W_0$
        and a target outside $\vis(w)$. Then $C$ holds.
    \end{conditions}
\end{theorem}

\begin{proof}
    \emph{Signs.} The first part of \condref[]{plucker} follows from the
    polynomial Grassmann--Pl\"ucker identity
    \begin{multline*}
        \orient(p, g, b)\orient(p, c, d) - \orient(p, g, c)\orient(p, b,
        d)\\ + \orient(p, g, d)\orient(p, b, c) = 0
    \end{multline*}
    (see Knuth~\cite{Knu92}), since three nonzero real numbers summing to
    zero cannot all share the same sign. The second part follows from the
    signed-area identity
    $\orient(p,q,r)=\orient(p,q,z)+\orient(q,r,z)+\orient(r,p,z)$. For
    \condref[]{radial}, put $d_q=\epsilon_q(q-p)$. Every $d_q$ points into
    the open half-plane strictly to the left of $pr$, and bilinearity
    gives $\sgn\det(d_q,d_s)=\epsilon_q\epsilon_s\chi(p,q,s)$. The order
    variables therefore record the strict counterclockwise order of the
    vectors $d_q$ within a single open half-plane, which is transitive.

    \emph{Hull case.} For \condref[]{hull-order}, traverse $\mV$
    counterclockwise: every interior target lies strictly to the left of
    each directed hull edge, and \propref{certified-structure} (i) places
    $l$, $\mH$, and $r$ consecutively when $\mH\ne\emptyset$. For
    \condref[]{hull-obstruction}, suppose a witness $w$ sees $s_1,s_2$
    and misses $v\in \mD$ with $v\in\inte\triangle(w,s_1,s_2)$. The
    witness $\missw{v}=w_{\mS\setminus\{v\}}$ also sees $s_1,s_2$, so
    \lemref{cone} gives $v\in\inte\triangle(\missw{v},s_1,s_2)$
    (immediately if $w=\missw{v}$). By \lemref{front-restoration}, $v$ is
    then a front vertex of $\mV$ from $\missw{v}$; since $\missw{v}$ sees
    both hull neighbors of $v$ and misses $v$, $v$ is certified,
    contradicting $v\in \mD$. For \condref[]{exit-edge} and
    \condref[]{certified-pocket}, \lemref{exit-edge} and
    \propref{certified-structure} (ii) supply the enclosing triangles,
    \lemref{hidden-front}, \lemref{side-corners}, and \lemref{separation}
    supply the neighbor and separation constraints, and
    \lemref{wedge-sep} gives wedge separation around the missed target in
    each family.

    \emph{Half-plane constraints.} Assign the auxiliary gate variable of
    each \condref[]{half-plane} instance to the truth value of its
    antecedent. When the antecedent holds, $w_1=w_\mS$ sees every target
    in $\mS$, so either $w_0$ sees $w_1$ or $e\in\vis(w_0)\cap\vis(w_1)\cap\mS$
    satisfies hypothesis~(iv) of \lemref{half-plane}, and
    \lemref{half-plane} yields all stated conclusions.

    \emph{Enclosure.} For \condref[]{paths}, suppose $u$ does not see $v$.
    Every path of links from $u$ to $v$ has at least two edges; if both
    open sides of $uv$ contained internal vertices of such paths,
    \lemref{enclosure} (ii) would make $uv$ a link. Select an open side
    of $uv$ containing no such path, so $u$ and $v$ lie in different
    connected components of the link subgraph induced on $u$, $v$, and
    the points of $\mX$ on that side; color the component of $u$ with one
    color and all remaining components with the opposite color. For
    \condref[]{triangles}, an auxiliary gate records whether all three
    sides are links, in which case \lemref{enclosure} (i) and convexity
    yield the interior links. For \condref[]{darts}, auxiliary gates
    record whether $(w_1,g,w_2,b)$ is a convex quadrilateral or a reflex
    dart with a given inward vertex, and \lemref{four-cycle-enclosure}
    gives the forced links. Finally, for \condref[]{cuts}, if a true
    realization falsified $C$, its Euclidean signs would match all fixed
    signs used in the derivation, so \lemref{iterated-enclosure} would
    force some $w\in\W_0$ to see a target outside $\vis(w)$, a
    contradiction.
\end{proof}

Seventy-three of the $74$ formulas consist solely of clauses generated
directly from \condref[]{plucker}--\condref[]{darts}. The remaining case,
\texttt{s2-d5-h0-00}, additionally includes saved cut clauses from
\condref[]{cuts}; during replay, each saved cut is re-verified directly
from its negated literals via \lemref{iterated-enclosure}, without
relying on any SAT solver output.

\section{Proof of the main theorem}
\seclab{proof-main}
\begin{theorem}
    \thmlab{main-six}
    Every simple polygon $P\subset\R^2$ satisfies $\vcdim(P)\le 6$.
    Conversely, as shown by Valtr~\cite{Val98}, some simple polygon
    shatters six targets in strictly convex position. Consequently,
    \[
        \max_P \vcdim(P)=\max_P \vcdimconv(P)=6,
    \]
    where both maxima range over all simple polygons $P\subset\R^2$.
\end{theorem}

\begin{proof}
    Suppose for contradiction that a simple polygon shatters at least
    seven targets. By \obsref{monotonicity} it shatters a seven-target
    subset, and \lemref{general-position} yields a general-position
    realization $(P,\mS,\W)$ with $\cardin{\mS}=7$. The classification of
    \secref{case-enumeration} assigns $(P,\mS,\W)$ to one of the $74$
    hull cases.

    By \thmref{chirotope-sound}, the Euclidean chirotope signs and true
    visibility links of $(P,\mS,\W)$, together with the constructed truth
    values for the auxiliary variables, satisfy every clause in the
    propositional formula of that hull case. The formula would therefore
    be satisfiable, contradicting its verified DRAT refutation
    (\secref{certificates}).

    For the matching lower bound, Valtr~\cite{Val98} constructed a simple
    polygon shattering six targets in strictly convex position, so six is
    attained even under the convex-position restriction.
\end{proof}

\subsection{Independent proof certificates}
\seclab{certificates}

\paragraph{The repository.}
All supplementary material is archived at
\url{https://github.com/sarielhp/vc_vis_6}. Its directory
\texttt{certificates/} contains the Julia formula generator, a manifest
listing the $74$ case identifiers along with the variable count, clause
count, and SHA-256 digest of each DIMACS formula, one DRAT refutation per
case, and the saved cut clauses for \texttt{s2-d5-h0-00}.

\paragraph{Replaying the refutations.}
The replay program first verifies that the $74$ case identifiers in the
manifest match the exhaustive enumeration of \secref{case-enumeration}.
For each case it regenerates the DIMACS formula, checks every saved cut
clause directly from \lemref{iterated-enclosure} by verifying that its
negated literals force some $w_\mT\in\W_0$ to see a target outside $\mT$,
confirms the SHA-256 digest against the manifest, and invokes the bundled
\texttt{drat-trim} checker of Wetzler \etal~\cite{WHH14} on the saved
DRAT refutation. A successful check prints one \texttt{verified} line per
case. With Julia, QuickEnv.jl, and a C compiler installed, running
\texttt{sh replay.sh} at the root of the repository replays all $74$
cases end to end.

\paragraph{Lean formalization.}
An independent Lean~4 development in \texttt{lean/} verifies the upper
bound formally. Its main theorem, \texttt{simple\_polygon\_vcdim\_le\_six},
proves that no configuration of seven or more targets and witnesses
satisfies the predicate \texttt{ShatteredSimplePolygon}, an axiomatic
interface bundling general position, open witness neighborhoods, and
polygonal enclosure. \lemref{general-position} together with the
enclosure lemmas of \secref{enclosure-cone} and \secref{enclosure-links}
discharges this interface for ordinary simple polygons in $\R^2$.

The Lean proof follows an earlier organization of the $74$-case split:
formalized geometric lemmas directly rule out the all-hull case $(7,0)$
and the $11$ cases in which the exit edges of consecutive certified
vertices violate counterclockwise order around $w_\mD$, while a RUP
checker proved sound in Lean refutes the remaining $62$ cases across
$1{,}596$ certificate files. Along the way, the formalization also
establishes an intermediate upper bound of $11$
(\texttt{simple\_polygon\_vcdim\_le\_eleven}, ruling out $12$ shattered
targets). The project builds with Lean~4.31.0 and Mathlib~v4.31.0; for
both theorems, \texttt{\#print axioms} reports only \texttt{propext},
\texttt{Classical.choice}, and \texttt{Quot.sound}, with no
\texttt{sorry} and no \texttt{native\_decide}.

\paragraph{Other supplementary material.}
An earlier encoder, run with the SAT solvers CaDiCaL and Glucose, also
refuted all $74$ cases; the proof in this paper does not depend on it.
The directory \texttt{earlier\_encoder/} preserves those encoders, solver
logs, and Z3 scripts that check algebraic subclaims.

\section{Conclusion}
\seclab{conclusion}
We have shown that no simple polygon shatters seven targets, matching
Valtr's six-target construction~\cite{Val98} and establishing that the
maximum visibility VC-dimension of a simple polygon is six, both in
general and for targets in strictly convex position. Our upper bound
reduces a seven-target configuration to $74$ convex-hull cases and
excludes each case by an independently checked propositional refutation
over triple orientations and visibility links. Can all $74$ cases---or
the upper bound $\vcdim(P)\le 6$ as a whole---be proved by a short,
purely geometric argument without computer-checked case refutations?

\BibTexMode{%

}%
\BibLatexMode{\printbibliography}

\appendix

\section{Proof of the general-position lemma}
\apndlab{general-position}

\begin{proofof}{\lemref{general-position}:}
    Fix witnesses $(w_\mT)_{\mT \subseteq \mS}$ in $P$ with $\vis_P(w_\mT) \cap \mS
    = \mT$. For every $\mT \subseteq \mS$ and every $s \in \mS \setminus \mT$, the
    segment $w_\mT s$ is not contained in $P$, so we may pick a point
    $z_{\mT, s} \in w_\mT s \setminus P$. Since $P$ is compact and $\mS$ is finite,
    \[
        \delta_1 = \min_{\mT \subseteq \mS,\; s \in \mS \setminus \mT}
        \dist(z_{\mT, s}, P) > 0 \qquad (\text{taking }\delta_1 = 1\text{
        if }\mS = \emptyset).
    \]
    Following the perturbation argument of Gilbers and Klein~\cite{GK14},
    choose a simple polygon $P'$ with $P\subset\inte P'$ inside the
    $(\delta_1/2)$-neighborhood of $P$. Concretely, after removing any
    redundant collinear vertices of $P$, shift the supporting line of each
    edge of $P$ outward by a common distance $\varepsilon > 0$ and join
    consecutive shifted lines at their intersection. As $\varepsilon \to
    0$, these intersections converge to the original vertices of $P$. For
    sufficiently small $\varepsilon > 0$, the shifted edges preserve their
    cyclic order and positive lengths, nonincident edges remain disjoint,
    and the resulting simple polygon $P'$ satisfies $P \subset \inte P'$
    while staying within distance $\delta_1/2$ of $P$. Consequently, every
    prescribed link $w_\mT s$ ($s \in \mT$) lies in $\inte P'$, whereas
    every witness point $z_{\mT,s}$ ($s \in \mS \setminus \mT$) remains in
    the open exterior $\R^2 \setminus P'$.

    Next, choose safe perturbation disks around the points. For each
    $s \in \mT$, the compact segment $w_\mT s \subset \inte P'$ has
    positive distance from $\partial P'$, and for each $s \in \mS \setminus
    \mT$, the point $z_{\mT, s}$ lies in the open set $\R^2 \setminus P'$.
    Moreover, if $\mS$ is in strictly convex position with $\cardin{\mS}
    \ge 2$, each $s \in \mS$ has positive distance from $\CHX{\mS \setminus
    \{s\}}$. By continuity, we can choose an open disk $D_p \subset \inte
    P'$ centered at each indexed point $p$ in the formal disjoint union
    $\mS \sqcup \Set{w_\mT}{\mT \subseteq \mS}$ (taking $D_{w_\mT} = D_s$
    whenever $w_\mT = s$ in $\R^2$) such that for all choices $s' \in D_s$
    and $w'_\mT \in D_{w_\mT}$:
    \begin{compactenumi}
    \item strict convex position of $\mS'$ holds whenever $\mS$ is in
        strictly convex position,
    \item $w'_\mT s' \subset \inte P'$ whenever $s \in \mT$ (using
        convexity of $D_s \subset \inte P'$ when $w_\mT = s$), and
    \item $w'_\mT s' \cap (\R^2 \setminus P') \neq \emptyset$ whenever
        $s \in \mS \setminus \mT$.
    \end{compactenumi}

    Finally, visit the $\cardin{\mS}+2^{\cardin{\mS}}$ indexed points $p$
    one by one and pick the perturbed point $p' \in D_p$ outside the
    finitely many points and lines spanned by the points already chosen.
    This yields \condref[]{general-position}, and shrinking each disk
    $D_{w_\mT}$ to a smaller open disk centered at the chosen witness
    $w'_\mT$ yields \condref[]{witness-disks}.
\end{proofof}

\end{document}